\documentclass[apj]{emulateapj}
\usepackage{apjfonts}
\usepackage{amsmath}
\usepackage{graphicx}
\usepackage{booktabs}
\shorttitle{New Numerical Scheme for Resistive RMHD}
\shortauthors{TAKAMOTO \& INOUE}

\begin{document}


\title{A New numerical scheme for resistive relativistic MHD 
       using method of characteristics}


\author{Makoto Takamoto}
\affil{Theoretical Astrophysics Group, 
Department of Physics, Kyoto University}
%\email{aastex-help@aas.org}

\author{Tsuyoshi Inoue}
\affil{Division of Theoretical Astronomy, 
       National Astronomical Observatory of Japan,}
%\affil{Department of Physics and Mathematics, Aoyama Gakuin University
%       }
%\email{aastex-help@aas.org}


%\altaffiltext{1}{Visiting Astronomer, Cerro Tololo Inter-American Observatory.
%CTIO is operated by AURA, Inc.\ under contract to the National Science
%\altaffiltext{2}{Society of Fellows, Harvard University.}
%\altaffiltext{3}{present address: Center for Astrophysics,
%    60 Garden Street, Cambridge, MA 02138}
%\altaffiltext{4}{Visiting Programmer, Space Telescope Science Institute}
%\altaffiltext{5}{Patron, Alonso's Bar and Grill}


\begin{abstract}
We present a new numerical method of special relativistic resistive magnetohydrodynamics 
with scalar resistivity 
%that can solve fluid energy dominant region accurately. 
that can treat a range of phenomena, 
from nonrelativistic to relativistic 
(shock, contact discontinuity, and Alfv\'en wave).
%This solver calculates the numerical flux of fluid by using approximate Riemann solver, 
The present scheme calculates the numerical flux of fluid by using an approximate Riemann solver, 
and electromagnetic field by using the method of characteristics. 
%Since this scheme use two characteristic velocities differently, 
Since this scheme uses appropriate characteristic velocities, 
it is capable of accurately solving problems 
%it can solve accurately problems 
%that can not approximate as magnetohydrodynamics 
that cannot be approximated as ideal magnetohydrodynamics 
%or whose characteristic velocity is much lower than light. 
and whose characteristic velocity is much lower than light velocity. 
%and whose characteristic velocity is much lower than speed of light. 
The numerical results show that 
our scheme can solve the above problems 
as well as nearly ideal MHD problems. 
%Especially, 
%our new scheme is suitable for systems with initially weak magnetic field, 
Our new scheme is particularly well suited to systems with initially weak magnetic field, 
and mixed phenomena of relativistic and non-relativistic velocity; 
for example, MRI in accretion disk, and super Alfv\'enic turbulence. 
%この論文では新たな特殊相対論的散逸磁気流体のschemeを発表する。
%このschemeは流体の数値流速は流体の近似リーマンsolverで、電磁場は特性曲線法によって解く。
%２つの異なる特性速度を用いて解くために、散逸がきいてMHD近似が成り立たない
%ような問題や、high densityで特性速度が光速よりも十分遅いような問題も精度よく
%解くことが出来る。
%数値テストの結果は既存のコードで精度よく解けた問題も同様に解くことが出来、
%先述のような問題も精度よく解くことが出来ることを示している。
%この新しいスキームは初期に磁場が弱い系やMRIの起こるaccretion diskなどの
%問題に応用するのに適している。
%\textcolor{blue}{
%In this paper, 
%we explain one-dimensional scheme, 
%since our new algorithm is not related to spatial-dimensions. 
%We will explain the multi-dimensional scheme in the next paper. 
%}
\end{abstract}





\section{Introduction}\label{sec:sec1}
The magnetohydrodynamics (MHD) approximation has some interesting properties, 
for example, the flux freezing and magnetic pressure; 
the former can be used for the collimation of the jet, 
and the latter for the acceleration of the plasma. 
Thus, 
the magnetic field is considered an essential ingredient 
for many astrophysical phenomena. 
In particular, 
many observations indicate that 
most of the high energy phenomena in astrophysics 
are related to the strongly magnetized relativistic plasma around some compact objects, 
for example, 
AGN~\citep{A93,UP95}, 
relativistic jet~\citep{BK79,MR99}, 
pulsar wind~\citep{RG74,C09}, 
gamma-ray bursts~\citep{W93,PIR04}, 
and so on.
Since it is extremely difficult to solve the relativistic MHD (RMHD) equations analytically, 
the theoretical investigations in fully nonlinear regimes 
are mainly based on the numerical simulations~\citep{MG04,IAI10}. 
Most of these studies approximates the plasma as the ideal RMHD fluid. 
One reason for this is that 
the ideal RMHD is an excellent approximation of high energy phenomena 
for ordinary parameters. 
However, 
when one considers extreme phenomena, 
such as the neutron star mergers, or the central engines of GRB, 
the electrical conductivity can be small, 
and highly resistive regions may appear. 
In addition, 
when one considers the magnetic reconnection, 
the resistivity plays an essential role in this phenomenon. 
Magnetic reconnection is one of the most important phenomena, 
since it is highly dynamic, 
and it changes magnetic field energy into fluid energy
~\citep{ZY09,Z09,L07}. 
Though numerical results of ideal RMHD exhibit magnetic reconnection, 
this originates in the purely numerical resistivity, and this is unphysical. 
For this reason, 
using resistive RMHD is important for the understanding of reconnection 
and related phenomena. 
%Besides, 
%observation of solar activity and the numerical results of PIC simulation indicate that 
%the magnetic reconnection is driven by the anomalous resistivity in the current sheet 
%that is anomalously higher than that in other region
%~\citep{Petschek(1964),Coppi & Friedland(1971),Smith & Priest(1972)}.
%For these reasons, 
%one has to use resistive RMHD solver 
%when one consider problems including magnetic reconnection. 
%流体は磁場に凍り付くためjetのcollimationに関係すると考えられ、
%さらに磁気圧の勾配力により流体を加速する。 
%宇宙の様々な現象には磁場が関係していると考えられている。
%これらのような高エネルギー天体現象の解明のために様々なideal RMHD solverが考案されている。
%相対論的なプラズマは通常オーム散逸は極めて小さいと考えられている。
%$\sigma \sim 10^{24} (10^9 K / T)^2 (\rho /10^{14} g cm^{-3})^{3/4} s^{-1}$
%より、高温で低密度の場合は磁気散逸がきいてくる。
%更に非熱的な高エネルギー現象を起こす過程として重要なreconnectionがあるが、
%この現象の原因となるcurrent sheetでは異常抵抗という、通常よりも大きな抵抗が働くことが
%示唆されている。
%ideal MHDで起きるreconnectionは数値散逸の結果であるので非物理的である。
%このためresistive RMHDを取り扱う必要がある。

In order to consider Ohmic dissipation, 
one only has to take into account an additional term $- \nabla \times (\nabla \times {\bf B}) / \sigma$ 
in the induction equation 
of non-relativistic MHD. 
However, 
similar to other non-relativistic dissipation, 
%for example, viscosity and thermal conduction, 
this induction equation is parabolic 
and it is well-known that 
this equation is acausal. 
As a result, 
if one takes into account Ohmic dissipation in a relativistic MHD 
in a similar way, 
the equation inevitably includes unphysical exponential growing modes, 
and unstable for small perturbations 
similar to other dissipation~\citep{HL83,HL85}. 
This unphysical divergence results from the fact that 
one neglects the time derivative of the electric field 
in the induction equation with Ohmic dissipation. 
For this reason, 
when one takes into account the Ohmic dissipation, 
one has to consider the time evolution of the electric field, 
that is, 
one has to deal with the relativistic electromagnetic hydrodynamic equation. 
This equation is a telegrapher equation, 
and satisfies the causality. 
%In the following, 
%we explane such causal and stable numerical scheme of relativistic resistive MHD. 
%因果律について記述
%
%非相対論でオーム散逸を考える場合、ideal MHD方程式の磁場の発展方程式に
%付加項として$\nabla \times (\nabla \times B / \sigma)$
%を加えるだけでよい。
%しかしこの式は非相対論での粘性や熱伝導と同じく、方程式系は放物型になり、
%よく知られているように因果律を破ってしまう。
%この結果、相対論的な方程式で同様の方法でオーム散逸を考慮した場合、
%他の散逸と同様に因果律を破ることが原因となって線形摂動に指数関数的に発散する非物理的なモードが必ず入ってきてしまう。
%この因果律の破れは磁場の発展方程式にオーム散逸を考慮する際に、
%電場の時間微分を無視したことに起因する。
%このことにより、相対論的MHDでオーム散逸を因果律を守るように考慮するには、
%電場の発展を考慮した相対論的EMHDを考える必要がある。
%このようにすることでRRMHDは電信方程式になり、因果律を守るのである。
%この論文では以降でこのような因果律を守り安定な数値スキームを説明する。

In this paper, 
we present a new numerical scheme for the resistive RMHD. 
There are several examples of pioneering work for resistive RMHD, 
for example, 
Komissarov (2007, hereafter K07) proposed numerical method 
that solves hyperbolic fluxes by using the Harten-Lax-van Leer (HLL) prescription, 
and damping of the electric field by Ohmic dissipation that is very stiff 
by using Strang-splitting techniques; 
Palenzuela et al. (2009, hereafter P09) proposed a numerical method 
that solves hyperbolic fluxes by Local Lax-Friedrichs approximate Riemann solver, 
and the stiff part by using implicit-explicit (IMEX) Runge Kutta methods. 
However, 
these methods use light velocity as the characteristic velocity, 
and their numerical solutions are diffusive when one considers problems 
%and they are incapable of providing accurate solutions for problems  
whose characteristic velocity is much lower than light velocity. 
This indicates that 
%they cannot solve accurately many important high plasma $\beta$ problems, 
their numerical solutions are diffusive in many important high plasma $\beta$ dynamics, 
and also their solutions become highly diffusive 
when the characteristic velocity of phenomena is much lower than light velocity. 
In particular, 
when one solves the dynamics of the accretion disk around a black hole 
with a relativistic jet, 
one has to use relativistic resistive MHD code 
that can solve both highly relativistic and non-relativistic dynamics with resistivity 
for the following three reasons: 
(1) the saturation of the magnetorotational instability (MRI) depends on the resistivity; 
(2) the dynamics of an accretion disk are not ordinarily relativistic, 
especially, 
the dynamics of the MRI is sub-Alfv\'enic; 
(3) the dynamics of the jet are highly relativistic. 
For these reasons, 
previous schemes are diffusive in such phenomena, 
%previous schemes cannot deal with such phenomena, 
and we need more accurate numerical schemes. 
%such as magnetorotational instability (MRI) in accretion disk around black hole, 
%Though there are some resistive RMHD schemes(refs), 
%our scheme can solve problems accurately 
%whose characteristic velocity of fluid and electromagnetic field are quit different from speed of light, 
%for example, high plasma $\beta$ problems 
%that is important when one considers magnetorotational instability (MRI) in accretion disk around black hole
We are developing a new numerical scheme 
capable of accurately solving problems 
whose characteristic velocity is quite different from light velocity. 
In this scheme, 
we obtain numerical flux of fluid by using sound velocity as the characteristic velocity, 
and numerical flux of electromagnetic field by using appropriate characteristic velocities of RMHD. 
%and numerical flux of electromagnetic field by using fast-magnetic wave velocity as the characteristic velocity. 
%and numerical flux of electromagnetic field by using Alfv\'en velocity as the characteristic velocity. 
This enables us to obtain accurate numerical results 
when we consider problems 
whose characteristic velocity is much lower than light velocity. 
In addition, 
P09 pointed out that 
the Strang-splitting method used in the Komissarov method is unstable 
when applied to discontinuous flows with large conductivities. 
However, 
we find that 
this problem is not related to the Strang-splitting method, 
but the evolution of electric field ${\bf E}$ during the primitive recovery, 
that is introduced in the method by P09. 
By considering this procedure, 
we can apply the Strang-splitting method 
to discontinuous flows with large conductivities. 
%resistive RMHDのschemeとしては、これまでに
%Komissarovがpionier workとしてHLL + strang's splitting methodによるコードを開発しており、 
%またPalenzuela et al.はKomissarov codeで解けなかった不連続が存在するhigh \sigmaの問題を解くことが出来る
%Local Lax-Friedrichs approximate Riemann solver + implicit-explicit (IMEX) Runge Kutta methods
%というコードを開発している
%ところがこれらのコードはそれぞれ特性速度として光速、もしくは系の最も早い特徴的速度を用いてしまっているため、
%音速とアルフヴェン速度が大きく異なる問題や、音速やアルフヴェン速度が光速よりも大きく小さい場合に精度が落ちてしまう。
%これらのために過去のコードでは、
%ブラックホール周りの降着円盤のMRIなどのhigh betaの非常に重要な問題が精度よく解けないと考えられる。

%新たに特性速度が光速よりも大きく小さい、もしくは特性速度が電磁場と流体で大きく
%ことなるような場合も精度よく解けるschemeを開発した。
%resistive RMHDは電場の発展を考慮した電磁流体方程式を解くことになるが、
%私たちは流体部分は音速を、電磁気部分はfast magnetosonic速度を特性速度にしてfluxを計算することにより
%特性速度が光速から大きく異なる場合にも精度よく計算することが出来る。
%またPalenzuela et al.はStrang splitting methodは不連続面がありhigh sigmaの場合は解くことが出来ない
%としていたが、我々はこれはSplitting methodを用いるかimplicitを用いるかには関係なく、
%Palenzuela et al.で考慮されていたprimitive recoveryの際に電場の発展にあることを発見した。
%これを組み込むことにより、我々はsplitting methodを用いなおかつ不連続面のあるhigh sigmaの問題も
%精度よく解くことが出来る。

%\textcolor{blue}{
%Since the advantage of our method is not related to spatial-dimension, 
%we present only one-dimensional explanation in this paper. 
%The multi-dimensional explanation will be presented in next paper.
%}

This paper is organized as follows. 
In Section \ref{sec:sec2}, 
the basic equations of resistive RMHD are presented. 
In Section \ref{sec:sec3}, 
we present the numerical method. 
Results of numerical test problems previously presented are shown in Section \ref{sec:sec4}. 
In Section \ref{sec:sec5}, 
we present results of numerical test problems 
that cannot be solved accurately by previous codes. 
\section{Basic Equations}\label{sec:sec2}
Throughout this paper, we use the units
\begin{equation}
c = 1
,
\end{equation}
In Cartesian coordinates, the Minkowski metric tensor $\eta_{\mu \nu}$ is given by
\begin{equation}
\eta_{\mu \nu} = \mathrm{diag}(- 1, 1, 1, 1)
.
\end{equation}
Variables indicated by Greek letters take values from $0$ to $3$, 
and those indicated by Roman letters take values from $1$ to $3$.

\subsection{The Maxwell equations}\label{sec:sec2.1}
The covariant Maxwell equations can be written as 
\begin{eqnarray}
\partial_{\nu} F^{\mu \nu} &=& I^{\mu}
,
\\
\partial_{\nu} ^*F^{\mu \nu} &=& 0
,
\end{eqnarray}
where $F^{\mu \nu}$ is the Maxwell tensor, 
$^*F^{\mu \nu}$ the Faraday tensor, 
and $I^{\mu}$ the four-vector of electric current. 

If we consider highly ionized plasma, 
the electric and magnetic susceptibilities can be neglected. 
Then, one has 
\begin{equation}
^*F^{\mu \nu} = \frac{1}{2} e^{\mu \nu \rho \sigma} F_{\rho \sigma}
,
\end{equation}
where 
\begin{equation}
e^{\mu \nu \rho \sigma} = \sqrt{-g} \epsilon_{\mu \nu \rho \sigma}
,
\end{equation}
is the Levi-Civita alternating tensor of space-time, 
and $\epsilon_{\mu \nu \rho \sigma}$ is the four-dimensional Levi-Civita symbol. 

We introduce a future-directed unit timelike vector $n^{\mu}$ 
normal to a spacelike hypersurface $\Sigma$. 
Using $n^{\mu}$, we can decompose the Maxwell tensor 
into following forms: 
\begin{equation}
F^{\mu \nu} = n^{\mu} E^{\nu} - n^{\nu} E^{\mu} 
+ n_{\rho} e^{\rho \mu \nu \sigma} B_{\sigma}
\label{dec_M}
.
\end{equation}
Similarly, the current four-vector $I^{\mu}$ can be decomposed into: 
\begin{equation}
I^{\mu} = q n^{\mu} + J^{\mu}
\label{dec_I}
,
\end{equation}
where $q$ is the charge density observed in the rest frame of $n^{\mu}$, 
and $J^{\mu}$ the conduction current satisfying $J^{\mu} n_{\mu} = 0$.
In the following, 
we consider only Minkowski spacetime, 
so $n^{\mu} = (1, 0, 0, 0)$.

By using the decomposition of the Maxwell tensor Eq. (\ref{dec_M}) and 
the current four-vector (\ref{dec_I}), 
the Maxwell equations can be split into the familiar set 
\begin{eqnarray}
\nabla \cdot \mathbf{E} &=& q
\label{Maxwell_1}
,
\\
\nabla \cdot \mathbf{B} &=& 0
\label{Maxwell_2}
,
\\
\partial_t \mathbf{E} - \nabla \times \mathbf{B} &=& - \mathbf{J}
\label{Maxwell_3}
,
\\
\partial_t \mathbf{B} + \nabla \times \mathbf{E} &=& \mathbf{0}
\label{Maxwell_4}
.
\end{eqnarray}
From Maxwell equations, 
we can derive the electric charge conservation law 
\begin{equation}
\partial_t q + \nabla \cdot \mathbf{J} = 0
.
\end{equation}

\subsection{The hydrodynamic equations}\label{sec:sec2.2}
The relativistic hydrodynamic equations can be obtained from the conservation of 
mass, momentum, and energy 
\begin{eqnarray}
\partial_{\mu} N^{\mu} &=& 0
,
\\
\partial_{\nu} T^{\mu \nu} &=& 0
,
\end{eqnarray}
where $N^{\mu}$ is the mass density current and 
$T^{\mu \nu}$ the energy-momentum tensor defined respectively as
\begin{eqnarray}
N^{\mu} &=& \rho u^{\mu}
,
\\
T^{\mu \nu} &=& T^{\mu \nu}_{\mathrm{fluid}} + T^{\mu \nu}_{\mathrm{EM}}
,
\end{eqnarray}
where
\begin{eqnarray}
T^{\mu \nu}_{\mathrm{fluid}} &\equiv& \rho h u^{\mu} u^{\nu} + p \eta^{\mu \nu}
,
\\
T^{\mu \nu}_{\mathrm{EM}} &\equiv& F^{\mu \rho} F^{\nu}_{\rho} 
- \frac{1}{4} (F^{\rho \sigma} F_{\rho \sigma}) \eta^{\mu \nu}
.
\end{eqnarray}
Here $h = 1 + \epsilon + p / \rho$ is the specific enthalpy, 
$\rho$ is the proper rest mass density, 
$p$ is the thermodynamic pressure, 
and $\epsilon$ is the specific internal energy.

The evolution equation of a relativistic resistive MHD is
\begin{eqnarray}
\partial_t
\left(
 \begin{array}{c}
   D \\
   m^i \\
   e
 \end{array}
\right)
+ \partial_j
\left(
 \begin{array}{c}
   F_D^j \\
   F_m^{ij} \\
   F_e^j
 \end{array}
\right)
= 0,
\label{fluid}
\end{eqnarray}
where $D$, $m^i$, $e$ is the density, momentum density, total energy density. 
In the laboratory frame, $D$, ${\bf m}$, $e$ are given by
\begin{eqnarray}
D &=& \gamma \rho
\label{D}
, \\
\mathbf{m} &=& \rho h \gamma^2 {\bf v} + \mathbf{E \times B}
\label{m}
, \\
e &=& \rho h \gamma^2 - p + \frac{1}{2} (E^2 + B^2)
\label{e}
,
\end{eqnarray}
where ${\bf v}$ is the fluid three-velocity, $\gamma = (1 - v^2)^{-1/2}$ is the Lorentz factor, 
and numerical fluxes are
\begin{eqnarray}
F_D^i &=& D v^i 
,
\\
F_m^{ij} &=& m^i v^j + p \eta^{ij}
- E^i E^j - B^i B^j 
+ \frac{1}{2} (E^2 + B^2) \eta^{ij}
,
\\
F_e^i &=& m^i
.
\end{eqnarray}

This is the most common form of perfect fluid equations for the numerical hydrodynamics.

\subsection{Ohm's law}\label{sec:sec2.3}
The system of Eqs. (\ref{Maxwell_1}) - (\ref{Maxwell_4}), (\ref{fluid}) 
is closed by means of Ohm's law. 
Although there are various forms of Ohm's law, 
we consider only the simplest kind of relativistic Ohm's law 
that accounts only for the plasma resistivity, and 
that assumes that it is isotropic 
similar to previous studies K07 and P09.
In the covariant form, it is given by
\begin{equation}
I^{\mu} = \sigma F^{\mu \nu} u_{\nu} + q_0 u^{\mu}
\label{Ohm_cov}
,
\end{equation}
where $\sigma = 1 / \eta$ is the conductivity, 
$\eta$ is the resistivity, 
and $q_0 = - I_{\mu} u^{\mu}$ is the electric charge density as measured in the fluid frame. 

As the Maxwell equations and fluid equations, 
we can decompose Eq. (\ref{Ohm_cov}) into 3 + 1 form, 
and then the space component of Eq. (\ref{Ohm_cov}) is given by 
\begin{equation}
\mathbf{J} = \sigma \gamma [\mathbf{E} + \mathbf{v \times B} - (\mathbf{E \cdot v}) \mathbf{v}] 
+ q \mathbf{v}
\label{Ohm_lab}
,
\end{equation}
In the fluid rest frame, 
Eq. (\ref{Ohm_lab}) becomes 
\begin{equation}
\mathbf{J} = \sigma \mathbf{E}
.
\end{equation}

The ideal MHD limit of Ohm's law can be obtained 
in the limit of infinite conductivity ($\sigma \rightarrow \infty$). 
In this limit, 
Eq. (\ref{Ohm_lab}) reduces to
\begin{equation}
\mathbf{E + v \times B - (E \cdot v) v = 0}
.
\end{equation}
Splitting this equation into the components 
that are normal and parallel to the velocity vector, 
it becomes
\begin{eqnarray}
\mathbf{E}_{\perp} + \mathbf{v \times B} &=& \mathbf{0}
,
\\
\mathbf{E}_{\parallel} - \mathbf{(E \cdot v) v} &=& \mathbf{0}
,
\end{eqnarray}
From these equations, 
we can obtain the usual result 
\begin{equation}
\mathbf{E = - v \times B}
.
\end{equation}

%\begin{equation}
%- \partial_t \mathbf{E} = \mathbf{J}
%,
%\end{equation}
%
%\begin{eqnarray}
%\partial_t \mathbf{E}_{\parallel} &+& 
%\sigma \gamma [\mathbf{E}_{\parallel} - \mathbf{(E \cdot v) v}] = \mathbf{0}
%,
%\\
%\partial_t \mathbf{E}_{\perp} &+& 
%\sigma \gamma [\mathbf{E}_{\perp} - \mathbf{v \times B} = \mathbf{0}
%,
%\end{eqnarray}
%
%\begin{eqnarray}
%\mathbf{E}_{\parallel} &=& \mathbf{E}^0_{\parallel} \exp \left[ - \frac{\sigma}{\gamma} t \right]
%,
%\\
%\mathbf{E}_{\perp} &=& \mathbf{E}_{\perp}^* 
%+ (\mathbf{E}^0_{\perp} - \mathbf{E}^*_{\perp}) \exp \left[ - \sigma \gamma t \right]
%\end{eqnarray}
%where $\mathbf{E}_{\perp}^* = - \mathbf{v \times B}$
\section{Numerical Method}\label{sec:sec3}
In this section, 
we present our new numerical scheme for the resistive RMHD. 
Since the pioneering studies of resistive RMHD K07 and P09
use light velocity as the characteristic velocity, 
%they cannot solve problems accurately  
their solution becomes highly diffusive 
when characteristic velocity is much lower than light velocity. 
%This indicates that 
%they cannot solve accurately many important high plasma $\beta$ problems, 
%such as magnetorotational instability (MRI) in accretion disk around black hole. 
In our new scheme, 
we obtain numerical flux of fluid by using sound velocity as the characteristic velocity, 
%and numerical flux of electromagnetic field by using fast-magnetic wave velocity as the characteristic velocity. 
and numerical flux of the electromagnetic field by using Alfv\'en velocity as the characteristic velocity. 
This enables us to obtain accurate numerical results 
even when characteristic velocity is much lower than light velocity. 
%In the following sections, 
%we consider only the one-dimensional case. 
%We explain the multi-dimensional case in Sec. \ref{sec:sec3.6}. 
%We will explain the multi-dimensional case in our next paper. 
In the following sections, 
we consider the one-dimensional case. 
The extension to the multi-dimensional scheme 
using the constrained transport method 
~\citep{EH88,SN92}.
will be shown in our next paper. 

\subsection{Strang Splitting method}\label{sec:sec3.1}
The relativistic resistive MHD is hyperbolic-relaxation equations. 
In previous work K07 and P09, 
they assume that characteristic velocity is the speed of light. 
Thus, their schemes are highly diffusive 
when the characteristic velocity is lower than light velocity. 
For this reason, 
we apply the Strang splitting method~\citep{S68} 
and solve the basic equations by using each appropriate characteristic velocity. 

First, we split fluid equations Eq. (\ref{fluid}) as follows: 
\begin{eqnarray}
\partial_t
\left(
 \begin{array}{c}
   D \\
   m^i \\
   e
%   D \\
%   m_{fluid}^i +  m^i_{EM} \\
%   e_{fluid} + E_{EM}
 \end{array}
\right)
+ \partial_x
\left(
 \begin{array}{c}
   F_D^x \\
   F_{m,fluid}^{ix} \\
   F_{e,fluid}^x 
 \end{array}
\right)
+ \partial_x
\left(
 \begin{array}{c}
   0 \\
   F_{m,EM}^{ix} \\
   F_{e,EM}^x
 \end{array}
\right)
= 0,
\label{RRMHD}
\end{eqnarray}
%\begin{eqnarray}
%D &=& \gamma \rho
%, \\
%\mathbf{m}_{fluid} &=& \rho h \gamma^2 {\bf v}
%,
%\\
%\mathbf{m}_{EM} &=& \rho h \gamma^2 {\bf v} + \mathbf{E \times B}
%, 
%\\
%e_{fluid} &=& \rho h \gamma^2 - p + \frac{1}{2} (E^2 + B^2)
%,\\
%e_{EM} &=& \rho h \gamma^2 - p + \frac{1}{2} (E^2 + B^2)
%,
%\end{eqnarray}
where
\begin{eqnarray}
F_D^x &=& D v^x 
\label{flux_fluid_D}
,
\\
F_{m,fluid}^{ix} &=& m^i v^x + p \eta^{ix}
\label{flux_fluid_m}
,
\\
F_{e,fluid}^x &=& \rho h \gamma^2 v^x
\label{flux_fluid_E}
,
\\
F_{m,EM}^{ix} &=& - E^i E^x - B^i B^x 
+ \left[ \frac{1}{2} (E^2 + B^2) \right] \eta^{ix}
\label{flux_EM_m}
,
\\
F_{e,EM}^x &=& (\mathbf{E \times B})^x
\label{flux_EM_E}
. 
\end{eqnarray}
The flux of the fluid component $F^x_{fluid}$ can be calculated by using the Riemann solver; 
the flux of the electromagnetic component $F^x_{EM}$ can be calculated 
%by using $\mathbf{e}$ and $\mathbf{B}$ evolved by Maxwell equations.
by the method of characteristics. 

Next, 
we consider the Maxwell equations Eqs. (\ref{Maxwell_1}) - (\ref{Maxwell_4}). 
Eqs. (\ref{Maxwell_1}) and (\ref{Maxwell_2}) are not evolution equations 
but constraint equations, 
and we treat them separately from evolution equations. 
The evolution equations of $\mathbf{E}$ and $\mathbf{B}$ are 
Eqs. (\ref{Maxwell_3}) and (\ref{Maxwell_4}). 
By using Ohm's law Eq. (\ref{Ohm_lab}), 
Eq. (\ref{Maxwell_3}) reduces to 
\begin{equation}
\partial_t \mathbf{E} - \nabla \times \mathbf{B} = - 
\sigma \gamma [\mathbf{E} + \mathbf{v \times B} - (\mathbf{E \cdot v}) \mathbf{v}] 
- q \mathbf{v}
\label{Maxwell_3_2}
.
\end{equation}
The source term of this equation includes evolving variables $\mathbf{E}$, 
so this equation is a hyperbolic equation with stiff relaxation terms
that requires special care to capture the dynamics in a stable and accurate manner. 
Thus, we split the charge current $\mathbf{J}$ into two parts 
similar to K07
\begin{eqnarray}
\mathbf{J} &=& q \mathbf{v} + \mathbf{J}_c
\label{current}
,
\\
\mathbf{J}_c &=& 
\sigma \gamma [\mathbf{E} + \mathbf{v \times B} - (\mathbf{E \cdot v}) \mathbf{v}] 
\label{current_stiff}
.
\end{eqnarray}
Then, we split Eq. (\ref{Maxwell_3_2}) into two parts
\begin{eqnarray}
\partial_t \mathbf{E} - \nabla \times \mathbf{B} &=& - q \mathbf{v}
\label{Maxwell_3_3}
,
\\
\partial_t \mathbf{E} &=& - \mathbf{J}_c
\label{Maxwell_3_4}
.
\end{eqnarray}
Eq. (\ref{Maxwell_3_3}) is non-stiff equations, 
and Eq. (\ref{Maxwell_3_4}) is stiff equations. 

As a result, 
the evolution part of the Maxwell equations can be rewritten as 
\begin{eqnarray}
\partial_t \mathbf{B} + \nabla \times \mathbf{E} &=& \mathbf{0}
\label{Maxwell_4_2}
,
\\
\partial_t \mathbf{E} - \nabla \times \mathbf{B} &=& - q \mathbf{v}
\label{Maxwell_3_3_2}
,
\\
\partial_t \mathbf{E} &=& - \mathbf{J}_c
\label{Maxwell_3_4_2}
.
\end{eqnarray}
In component form, 
Eqs. (\ref{Maxwell_4_2}) and (\ref{Maxwell_3_3_2}) reduce to 
\begin{eqnarray}
\partial_t B^x &=& 0
\label{Maxwell_4_2_x}
,
\\
\partial_t B^y - \partial_x E^z &=& 0
\label{Maxwell_4_2_y}
,
\\
\partial_t B^z + \partial_x E^y &=& 0
\label{Maxwell_4_2_z}
,
\\
\partial_t E^x &=& - q v^x
\label{Maxwell_3_3_2_x}
,
\\
\partial_t E^y + \partial_x B^z &=& - q v^y
\label{Maxwell_3_3_2_y}
,
\\
\partial_t E^z - \partial_x B^y &=& - q v^z
\label{Maxwell_3_3_2_z}
.
\end{eqnarray}
We solve Eqs. (\ref{Maxwell_4_2_y}), (\ref{Maxwell_4_2_z}), 
(\ref{Maxwell_3_3_2_y}), and (\ref{Maxwell_3_3_2_z}) 
using method of characteristics (MOC), 
which will be shown in Sec. \ref{sec:sec3.2}. 
Eq. (\ref{Maxwell_3_3_2_x}) is solved using the Runge-Kutta method. 
The numerical scheme for the stiff equation Eq. (\ref{Maxwell_3_4_2}) 
will be shown in Sec. \ref{sec:sec3.3}. 
%Eqs. (\ref{Maxwell_4_2_x}) - (\ref{Maxwell_3_3_2_z}) are non-stiff equations, 
%and Eq. (\ref{Maxwell_3_3_2_x}) can be solved 
%using the ordinary differential equation solver. 
%For Eqs. (\ref{Maxwell_4_2_y}), (\ref{Maxwell_4_2_z}), 
%(\ref{Maxwell_3_3_2_y}), and (\ref{Maxwell_3_3_2_z}), 
%we use method of characteristics 
%that we will explain later. 
%We explain in Sec. \ref{sec:sec3.3} how to deal with stiff part Eq. (\ref{Maxwell_3_4_2}). 

\subsection{Method of characteristics}\label{sec:sec3.2}
The method of characteristics can be used to solve 
the initial value problems of advective and hyperbolic equations. 
As is well known, 
the Maxwell equations are hyperbolic, 
so we can solve the Maxwell equations accurately by using this method. 

The Maxwell equations for the transverse fields are 
Eqs. (\ref{Maxwell_4_2_y}), (\ref{Maxwell_4_2_z}), 
(\ref{Maxwell_3_3_2_y}), and (\ref{Maxwell_3_3_2_z}). 
By adding and subtracting these equations, for $E^y$, $B^z$, and $J^y$, 
we obtain 
\begin{eqnarray}
&&\left[ \partial_t \pm c_{ch} \partial_x \right] ^{\pm}F = - \frac{1}{2} J^y
\label{charact}
,
\\
&&^{\pm}F \equiv \frac{1}{2} (E^y \pm B^z)
,
\end{eqnarray}
where $c_{ch}$ is the characteristic velocity, and this is equal to the speed of light 
in ordinal Maxwell equations.

The transverse fields are recovered from $^{\pm}F$ by 
\begin{eqnarray}
E^y &=& ^+F + ^-F
\label{ch1}
,
\\
B^z &=& ^+F - ^-F
\label{ch2}
,
\end{eqnarray}
The left-hand side of Eq. (\ref{charact}) is the total derivative $dF / dt$ 
for an observer moving at velocity $\pm c_{ch}$. 

\begin{figure}[h]
% \centering
%  \includegraphics[width=7cm,clip]{.charact_Alfven_gmp.eps}
%  \includegraphics[width=7cm,clip]{./graphics/charact_Alfven_gmp.eps}
%  \includegraphics[width=7cm,clip]{./graphics/charact_Alfven_ch_gmp.eps}
%  \includegraphics[width=7cm,clip]{./charact_Alfven_ch_gmp.eps}
  \epsscale{1.0}
  \plotone{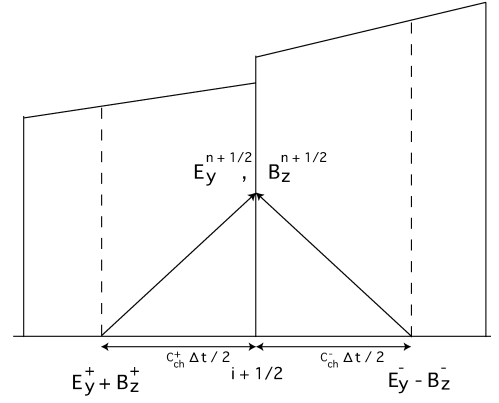}
%  \plottwo{charact_Alfven_ch_gmp.eps}{charact_Alfven2_ch_gmp.eps}
%  \includegraphics[width=7cm,clip]{./graphics/charact_Alfven.eps}
%  \includegraphics[width=7cm,clip]{./graphics/charact_fm.eps}
%  \includegraphics[width=7cm,clip]{./graphics/charact_light.eps}
%  \includegraphics[width=7cm,clip]{./charact_Alfven2_gmp.eps}
%  \includegraphics[width=7cm,clip]{./graphics/charact_Alfven2_gmp.eps}
%  \includegraphics[width=7cm,clip]{./graphics/charact_Alfven2_ch_gmp.eps}
%  \includegraphics[width=7cm,clip]{./charact_Alfven2_ch_gmp.eps}
  \plotone{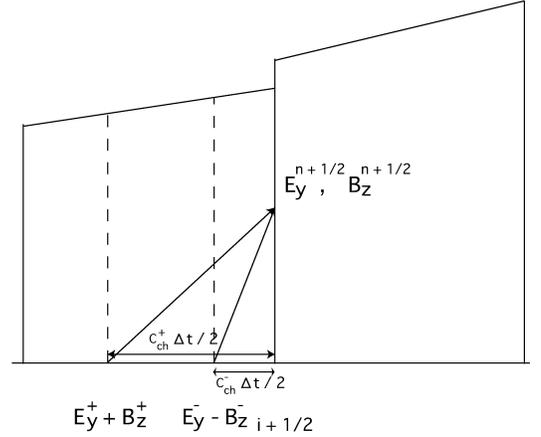}
%  \includegraphics[width=7cm,clip]{./graphics/charact_Alfven2.eps}
%  \includegraphics[width=7cm,clip]{./graphics/charact_fm2.eps}
%  \includegraphics[width=7cm,clip]{./graphics/charact_light.eps}
%  \caption{A schematic drawing of Eulerian-like characteristics of subsonic case 
  \caption{A schematic drawing of Eulerian-like characteristics 
           when one uses piecewise linear interpolation. 
           $c_{ch}$ is the characteristic velocity. 
           On the left is the subsonic case, 
           and on the right is the supersonic case. 
           These figures show that 
           half time-step transverse electromagnetic field $E^y$ and $B^z$
           are determined by the fields at the base of two characteristics.}
%  \caption{A schematic drawing of Eulerian-like characteristics of supersonic case 
%           when one uses piecewise linear interpolation. 
%           $c_A$ is the characteristic velocity.
%           This figure shows that 
%           half timestep transverse electromagnetic field $E^y$ and $B^z$
%           are determined by the fields at the base of two characteristics.
%           }
  \label{1}
%  \label{2}
\end{figure}

Let us consider conservative discretizations of Eqs. (\ref{Maxwell_4_2_z}) and (\ref{Maxwell_3_3_2_y}): 
\begin{eqnarray}
\bar{B}_{z,i}^{n+1}  &=& \bar{B}_{z,i}^{n} + \frac{\Delta t^n}{\Delta x_i} 
\left[ (E^y_{i+1/2})^{n+1/2} - (E^y_{i-1/2})^{n+1/2} \right]
\label{Maxwell_4_2_2_z}
,
\\
\bar{E}_i^{y,n+1}  &=& \bar{E}_{i}^{y,n} - \frac{\Delta t^n}{\Delta x_i} 
\left[ (B^z_{i+1/2})^{n+1/2} - (B^z_{i-1/2})^{n+1/2} \right] 
\nonumber
\\
&-& q^{n+1/2}_i v_i^{y,n+1/2} \Delta t^n
\label{Maxwell_3_3_3_y}
,
\end{eqnarray}
where superscript $n$ means the time-step, 
and subscript $i$ means the coordinate of cell center. 
Using Eqs. (\ref{ch1}) and (\ref{ch2}), 
we can obtain the numerical flux of 
Eqs. (\ref{Maxwell_4_2_2_z}) and (\ref{Maxwell_3_3_3_y}). 
(See Fig. \ref{1}.). 
The same procedure can be done for time advance of $E^z$, $B^y$. 

The characteristic velocity of the Maxwell equations in vacuum is light velocity. 
However, 
since we consider the electromagnetic hydrodynamics equations, 
appropriate characteristic velocity has to be used for them. 
Also, 
because we consider resistive systems, 
the characteristic velocity varies with the conductivity $\sigma$ and the scale of wave modes. 
For example, 
as shown in Appendix. %Sec. \ref{sec:seca2}, 
the transverse electromagnetic hydrodynamic waves propagate with the light velocity 
when $k / \sigma$ is large, 
where $k$ is the wave number, 
and they propagate with the Alfv\'en velocity 
when $k / \sigma$ is smaller than a critical value 
depending on $\rho, h$, and $|B|$. 
Because of the finite resistivity, 
the frequency of the transverse waves has an imaginary part $\omega_I$ (damping rate), 
which is a increasing function of $k / \sigma$. 
In this scheme, 
we use the light velocity as the characteristic velocity 
when $\sigma$ is smaller than the critical value; 
when $\sigma$ is larger than the critical value, 
we use appropriate magnetohydrodynamic characteristic velocities. 
The critical value is determined so that 
the transverse waves 
whose phase velocities are light velocity 
are dissipated within the numerical integration timestep $\Delta t$. 
%Note that 
%because of the finite resistivity, 
%the frequency of the transverse waves has an imaginary part $\omega_I$ (damping rate), 
%and the damping rate is a increasing function of $k / \sigma$. 
A detailed procedure to judge 
whether we use the light velocity or magnetohydrodynamic characteristic velocities 
is given in Appendix. %Sec. \ref{sec:seca2}. 

In addition to the numerical flux of the Maxwell equations, 
the characteristic velocity is also required to construct the Maxwell stress terms and the Poynting flux term. 
When the characteristic velocity obtained from the analysis of the transverse waves 
is the light velocity, 
we use the light velocity as the characteristic velocity for them; 
when the transverse wave characteristic velocity is the Alfv\'en velocity, 
we use the characteristic velocities for them as follows. 
Note that 
if the following characteristic velocities are not used, 
numerical integration becomes unstable, 
and unstable numerical oscillation occurs. 
%In ordinary Maxwell equations, 
%the characteristic velocity is speed of light. 
%However, 
%since we consider the electromagnetic hydrodynamics, 
%appropriate characteristic velocity of electromagnetic hydrodynamics has to be used. 
%In this scheme, 
%we obtain the characteristic velocity of Alfv\'enic mode 
%by perturbation of the full set of electromagnetic hydrodynamic equation 
%in fluid rest frame; 
%this characteristic velocity becomes speed of light 
%when conductivity $\sigma$ is small, 
%and becomes Alfv\'en velocity 
%when conductivity $\sigma$ is large. 
%We use speed of light as the characteristic velocity 
%if the obtained Alfv\'enic mode is speed of light, 
%and Alfv\'en velocity 
%if the mode is Alfv\'en velocity. 
%The detailed explanation is given in Sec. \ref{sec:seca2}. 
%
%If the Alfv\'enic mode is speed of light, 
%we use speed of light as the characteristic velocity 
%for obtaining numerical flux of Maxwell equation and electromagnetic hydrodynamic equation. 
%
%%since we are interested in MHD, 
%%so we use Alfv\'en velocity as characteristic velocity. 
%%so we use fast magnetosonic wave velocity as characteristic velocity. 
%In contrast, 
%if the Alfv\'enic mode is Alfv\'en velocity, 
%we have to use appropriate characteristic velocity 
%for each terms of numerical flux 
%following the MOC scheme:
%~\citep{Stone & Norman1(1992),Stone & Norman2(1992),Stone & Norman3(1992),Hawley & Stone(1995)}. 
%We note that 
%if the following characteristic velocities are not used, 
%numerical integration becomes unstable, 
%and unstable numerical oscillation occurs.

Then, 
the necessary procedures are as follows:  
\begin{enumerate}
  \item For the numerical flux of Maxwell equation 
%  Eqs. (\ref{Maxwell_4_2_y}), (\ref{Maxwell_4_2_z}), (\ref{Maxwell_3_3_2_y}), 
%  and (\ref{Maxwell_3_3_2_z}), 
  Eqs. (\ref{Maxwell_4_2_2_z}) and (\ref{Maxwell_3_3_3_y}), 
  we use the Alfv\'en wave velocity in laboratory frame 
  because the information of transverse electromagnetic fields are transmitted by the Alfv\'en wave
  \footnote{
In the case of small $B^x$ limit, the Alfv\'en velocity in laboratory frame $v_{AL}$ 
becomes $v_x$.  We find that when this $v_x$ is also small, numerical oscillations 
occur and numerical integration becomes occasionally unstable.  We can prevent this 
purious oscillations, if we use the characteristic velocity as $v_{AL} \rightarrow 0$
when $v_{AL} < 0.1 |B| / \sqrt{\rho h + |B|^2}$.  In this case, our scheme for
solving the 
Maxwell equations becomes equivalent to the HLLE scheme.  Note that introduction of 
this modification does not change any results presented in this paper.} . 
  In relativistic MHD, 
  the Alfv\'en velocity in laboratory frame $v_{AL}$ can be obtained by solving~\citep{ANI90} 
  \begin{equation}
    H a^2 - B^2 = 0, 
    \label{charact_Alfven_L}
  \end{equation}
  where $H = \rho h + b^2$, $a = \gamma (v_{AL} - v^x)$, $B = b^x - v_{AL} b^0$, and 
  $b^{\mu}$ is the covariant magnetic field defined as
  \begin{equation}
    b^{\mu} = \left[ \gamma \mathbf{v \cdot B}, 
    \frac{\mathbf{B}}{\gamma} + \gamma (\mathbf{v \cdot B}) \mathbf{v} \right]
    .
  \end{equation}

  \item For the Maxwell tension terms ${\bf - E E - B B}$ in Eq. (\ref{flux_EM_m}), 
  we use the Alfv\'en wave velocity in fluid comoving frame $v_{Ac}$
  because magnetic tension force is originated by the Alfv\'en wave. 
  In relativistic MHD, 
  the Alfv\'en velocity in fluid comoving frame is given by 
  \begin{equation}
    v_{Ac} = \frac{B^x}{\sqrt{H}}.  
    \label{charact_Alfven_R}
  \end{equation}

  \item For the Poynting flux ${\bf E \times B}$ of energy equation Eq. (\ref{flux_EM_E})
  and the Maxwell pressure terms $E^2 / 2  + B^2 / 2$ in Eq. (\ref{flux_EM_m}), 
  we use the fast magnetosonic wave velocity in laboratory frame 
  because the magnetic pressure originates in the magnetosonic wave. 
  In relativistic MHD, 
  fast magnetosonic wave velocity in the laboratory frame $v_{fm}$ can be obtained by solving
  \begin{eqnarray}
    \rho h (1 - c_s^2) a^4 = (1 - v_{fm}^2) [(|b|^2 + \rho h c_s^2) a^2 - c_s^2 B^2]
    \label{charact_fm}
  ,
  \end{eqnarray}
  Eq. (\ref{charact_fm}) is a quartic equation, 
  and in an ordinary one has to use the Newton-Raphson method or the quartic formula 
  for obtaining solutions. 
  However, 
  since our scheme splits the fluid part and the electromagnetic part, 
  the sound velocity $c_s$ can be set equal to zero. 
  Then, 
  the characteristic equation Eq. (\ref{charact_fm}) reduces to 
  \begin{equation}
    \rho h \gamma^2 (v^x - v_{fm})^2 = (1 - v_{fm}^2) | b |^2
    .
  \end{equation}
  By using the quadratic formula, 
  one can obtain solutions of above equation: 
  \begin{equation}
    v_{fm} = \frac{\rho h \gamma^2 v^x \pm |b| \sqrt{|b|^2 + (1 - (v^x)^2) \rho h \gamma^2} }
                   {\rho h \gamma^2 + |b|^2}
    \label{charact_fm_EM}
    .
  \end{equation}

\end{enumerate}

To sum up, 
we only have to substitute the appropriate characteristic velocities $v_{AL}, v_{Ac}$, and $v_{fm}$ 
into $c_{ch}$ in Eq. (\ref{charact}), 
and calculate the electromagnetic field $E, B$ at half time step. 
Then, 
the numerical fluxes of electromagnetic hydrodynamics equations are given by 
\begin{eqnarray}
F_{m,EM}^{ix} &=& - E_{Ac}^i E_{Ac}^x - B_{Ac}^i B_{Ac}^x 
+ \left[ \frac{1}{2} (E_{fm}^2 + B_{fm}^2) \right] \eta^{ix}
,
\\
F_{e,EM}^x &=& (\mathbf{E}_{fm} \times \mathbf{B}_{fm})^x
, 
\end{eqnarray}
where $E_{Ac}, B_{Ac}$ means that they are calculated by using the Alfv\'en velocity in comoving frame, 
and $E_{fm}, B_{fm}$ by using the fast magnetosonic wave velocity in laboratory frame. 
For the numerical flux of the Maxwell equation, 
one has to use the Alfv\'en velocity in laboratory frame $v_{AL}$ for the calculation. 
%Above conditions are necessary for the accurate and stable calculation in MHD approximation. 
%なお$\sigma$の小さい散逸の大きい場合はMHD近似が悪くなるため、
%fast magnetosonic velocityを特性速度にすると精度が落ちてしまう。
%このため私たちはAppendixで説明する方法で、
%考えている$\sigma$とresolutionの場合に特性速度を厳密に求め、
%それが光速の場合は特性速度を光速にして計算する。

%When we consider highly resistive case: $\sigma \gtrsim 0$, 
%we cannot use the MHD approximation; 
%%if we use fast magnetosonic velocity as the characteristic velocity, 
%if we use Alfv\'en velocity as the characteristic velocity 
%instead of light velocity, 
%our scheme becomes first-order in this case. 
%For this reason, 
%we calculate the exact characteristic velocity 
%using a method explained in Sec. \ref{sec:seca1}.
%This recovers spatial second-order of our scheme 
%in highly resistive case. 

%\subsection{\label{sec:sec3.3}Ohm's law}
\subsection{Stiff part}\label{sec:sec3.3}
As explained Sec. \ref{sec:sec3.1}, 
Eq. (\ref{Maxwell_3_4}) contains stiff terms. 
Following the previous work K07, 
we split the equation into components normal and parallel to the velocity vector. 
\begin{eqnarray}
\partial_t \mathbf{E}_{\parallel} &+& 
\sigma \gamma \left[ \mathbf{E}_{\parallel} - (\mathbf{E \cdot v}) \mathbf{v} \right] = 0
,
\\
\partial_t \mathbf{E}_{\perp} &+& 
\sigma \gamma \left[ \mathbf{E}_{\perp} + \mathbf{v \times B}) \right] = 0
,
\end{eqnarray}
Since we use the Strang splitting method, 
the right-hand side of the above equations can be considered constant other than the electric field $\mathbf{E}$. 
As a result, these equations can be solved analytically 
\begin{eqnarray}
\mathbf{E}_{\parallel} &=& 
\mathbf{E}^0_{\parallel} \exp \left[ - \frac{\sigma}{\gamma} t \right]
\label{exact1}
,
\\
\mathbf{E}_{\perp} &=& 
\mathbf{E}^*_{\perp} + ( \mathbf{E}^0_{\perp} -  \mathbf{E}^*_{\perp}) 
\exp \left[ - \sigma \gamma t \right]
\label{exact2}
,
\end{eqnarray}
where $E^*_{\perp} = - \mathbf{v \times B}$ 
and suffix 0 indicates the initial component. 
%In ordinary, 
If we use the explicit integrator, 
the stiff equation has to be solved in very small time steps $\Delta t$. 
However, since Eqs. (\ref{exact1}) and (\ref{exact2}) are formal solutions, 
we can avoid the stability constraints of the time step. 
In the context of ambipolar diffusion 
  in partially ionized plasma, 
a similar numerical technique 
  using the piecewise formal solution of stiff part 
    is known to be useful scheme
    ~\citep{IIK07,II08,II09}.

\subsection{Constraint Equations}\label{sec:sec3.4}
It is well known that 
Eqs. (\ref{Maxwell_1}) and (\ref{Maxwell_2}) are 
constraints on the Cauchy surface. 
Though Maxwell equations ensure that 
these constraints are preserved at all times, 
straightforward numerical integration of Maxwell equations does not preserve these properties 
because of the accumulated numerical error. 
This causes corruption of numerical results, 
and results in a crash in the end. 
For this reason, 
there are a number of numerical techniques 
for avoiding this problem. 
%For solving the constraint equations of the Maxwell equations 
%Eqs. (\ref{Maxwell_1}) and (\ref{Maxwell_2}), 
%and we adopt constrained transport (CT) for the magnetic field and 
We have implemented hyperbolic divergence cleaning for the electric field. 
The main idea of the hyperbolic divergence cleaning is that 
one defines new variable $\Psi$ as the deviation from constraint equations, 
and arranges a system of equations to decay or carry the deviation $\Psi$ out of the computational domain 
by high speed waves. 
%This will ensure the numerical deviations 
For the magnetic field, 
if one sets $B^x$ constant, 
the constraint equation can be satisfied in one-dimensional case. 
In the multi-dimensional case, 
we can implement constrained transport method~\citep{EH88,SN92}. 
The detailed implementation will be presented in our next paper. 
%(see also references~\citep{1992ApJS...80..753S,1992ApJS...80..791S,1992ApJS...80..819S,
%1995CoPhC..89..127H}). 
%}
%The detailed explanation of CT is presented in Sec. \ref{sec:sec3.6.2}
%(see also references~\citep{1992ApJS...80..753S,1992ApJS...80..791S,1992ApJS...80..819S,
%1995CoPhC..89..127H}). 
%In contrast, 

For hyperbolic divergence cleaning, 
we modify Eqs. (\ref{Maxwell_1}) and (\ref{Maxwell_3_3_2_x})
\begin{eqnarray}
&&\partial_t \Psi + \nabla \cdot \mathbf{E} = q - \kappa \Psi
, 
\label{DCE}
\\
&&\partial_t E^x + \partial_x \Psi = - q v^x
,
\end{eqnarray}
where $\Psi$ is a new dynamic variable 
and $\kappa$ a positive constant. 
Clearly, when we set $\Psi = 0$, 
we can recover standard Maxwell equation Eq. (\ref{Maxwell_1}). 
From these equations, 
we can obtain the telegrapher equation for $\Psi$
\begin{equation}
\partial_t^2 \Psi + \kappa \partial_t \Psi - \nabla^2 \Psi = 0
.
\end{equation}
Thus, $\Psi$ propagates at the speed of light 
and decays exponentially over a timescale $1 / \kappa$. 

Similar to Eq. (\ref{Maxwell_3_4}), 
Eq. (\ref{DCE}) contains stiff source terms. 
Thus, we split the equation into a stiff part and non-stiff part
\begin{eqnarray}
\partial_t \Psi + \nabla \cdot \mathbf{E} &=& q
,
\\
\partial_t \Psi &=& - \kappa \Psi
\label{DCE_stiff}
.
\end{eqnarray}
The analytical solution of Eq. (\ref{DCE_stiff}) is
\begin{equation}
\Psi = \Psi_0 \exp[- \kappa t]
\end{equation}
where $\Psi_0$ is the initial value of $\Psi$. 

\subsection{Primitive recovery}\label{sec:sec3.5}
In order to compute numerical flux 
(\ref{flux_fluid_D}), (\ref{flux_fluid_m}), (\ref{flux_fluid_E}), 
(\ref{flux_EM_m}), and (\ref{flux_EM_E}), 
the primitive variables $\{\rho, \mathbf{v}, p, \mathbf{B}, \mathbf{E} \}$ have to be 
recovered from the conserved variables $\{D, \mathbf{m}, e, \mathbf{B}, \mathbf{E} \}$. 
In conserved variables, 
$\mathbf{E}$ and $\mathbf{B}$ can be obtained by evolving the Maxwell equations. 
However, as pointed out by P09, 
it is more stable to perform evolution of stiff part 
Eqs. (\ref{exact1}) and (\ref{exact2}) 
during this primitive recovery process 
when $\sigma$ is large, 
i.e., ideal MHD approximation is valid. 
This is because when we consider MHD approximation, 
the electric field $\mathbf{E}$ is equal to $- \mathbf{v \times B}$; 
however, in general, primitive recovered $\mathbf{E}$ does not satisfy this relation. 
In what follows we explain the primitive recovery procedure following P09. 

\begin{enumerate}
\item Set an initial guess for the velocity by using previous time step value 
%$\mathbf{v} = \mathbf{v}^n$ when scheme is first-order 
%or $\mathbf{v} = \mathbf{v}^{n+1/2}$ when second-order. 
Then, evolve electric field $\mathbf{E}$ using Eqs. (\ref{exact1}) and (\ref{exact2}). 

\item Subtract Poynting flux and electromagnetic energy density from conserved variables, 
and new variables can be defined as follows:
  \begin{eqnarray}
    \mathbf{m}' &=& \rho h \gamma^2 {\bf v}
    , \\
    e' &=& \rho h \gamma^2 - p
    .    
  \end{eqnarray}
Then, variables $\{D, \mathbf{m}', e' \}$ are the ideal relativistic fluid conserved variables, 
and can be recovered by using the ordinary procedures. 

\item
Replace the initial guess for the velocity with the obtained velocity $\mathbf{v}$, 
and repeat the steps 1 - 3 until the primitive variables converge. 
%In this paper, 
%we iterate above procedure until the pressure $p$ converge. 
\end{enumerate}

\subsection{Algorithm}\label{sec:sec3.7}
In this section, 
we provide the detailed numerical algorithm. 

In Cartesian coordinates, 
the relativistic resistive MHD equations written in conservative fashion are 
\begin{eqnarray}
\partial_t
\left(
 \begin{array}{c}
   D \\
   m^i \\
   e
 \end{array}
  \right)
  + \partial_x
  \left(
 \begin{array}{c}
   F_D^x \\
   F_{m,fluid}^{ix} \\
   F_{e,fluid}^x 
 \end{array}
  \right)
  + \partial_x
  \left(
 \begin{array}{c}
   0 \\
   F_{m,EM}^{ix} \\
   F_{e,EM}^x
 \end{array}
  \right)
  = 0,
\end{eqnarray}
where
\begin{eqnarray}
D &=& \gamma \rho
, 
\\
\mathbf{m} &=& \rho h \gamma^2 {\bf v} + \mathbf{E \times B}
, 
\\
e &=& \rho h \gamma^2 - p + \frac{1}{2} (E^2 + B^2)
,
\\
F_D^x &=& D v^x 
,
\\
F_{m,fluid}^{ix} &=& m^i v^x + p \eta^{ix}
,
\\
F_{e,fluid}^x &=& \rho h \gamma^2 v^x
,
\\
F_{m,EM}^{ix} &=& - E^i E^x - B^i B^x 
+ \left[ \frac{1}{2} (E^2 + B^2) \right] \eta^{ix}
,
\\
F_{e,EM}^x &=& (\mathbf{E \times B})^x
. 
\end{eqnarray}

The electric field  $\mathbf{E}$ and magnetic field $\mathbf{B}$ are evolved by 
the Maxwell equations. 
If the Ohmic dissipation is considered, 
the Maxwell equations have stiff and non-stiff part. 
The non-stiff part is 
\begin{equation}
\partial_t U_{\mathrm{Maxwell}} + \partial_x F_{\mathrm{Maxwell}} = S_{\mathrm{non-stiff}}
,
\label{4_Maxwell}
\end{equation}
\begin{eqnarray}
U_{\mathrm{Maxwell}} &=& \left(
 \begin{array}{c}
   B^x \\
   B^y \\
   B^z \\
   E^x \\
   E^y \\
   E^z \\
   \Psi \\
   q
 \end{array}
\right)
,
F_{\mathrm{Maxwell}} = \left(
 \begin{array}{c}
   0 \\
   - E^z \\
   E^y \\
   \Psi \\
   B^z \\
   - B^y \\
   E^x  \\
   J^x
 \end{array}
\right)
,
\\
S_{\mathrm{non-stiff}} &=& \left(
 \begin{array}{c}
   0 \\
   0 \\
   0 \\
   - q v^x \\
   - q v^y \\
   - q v^z \\
   q  \\
   0
 \end{array}
\right)
,
\label{Maxwell_stiff}
\end{eqnarray}
where 
$J^x = \sigma [E^x + (\mathbf{v \times B})^x - (\mathbf{E \cdot v}) v^x] + q v^x$.
The Maxwell equations are consistent with the equation of charge conservation. 
However, numerical errors in general destroy the conservation law 
in a way similar to the constraint equations. 
Thus, the above equation contains the equation of charge conservation. 

As explained in Sec. \ref{sec:sec3.3} and \ref{sec:sec3.4}, 
the stiff-part is evolved by using the formal solution 
\begin{eqnarray}
\mathbf{E}_{\parallel} &=& 
\mathbf{E}^0_{\parallel} \exp \left[ - \frac{\sigma}{\gamma} t \right]
\label{exact_1.2}
,
\\
\mathbf{E}_{\perp} &=& 
\mathbf{E}^*_{\perp} + ( \mathbf{E}^0_{\perp} -  \mathbf{E}^*_{\perp}) 
\exp \left[ - \sigma \gamma t \right]
\label{exact_2.2}
,
\\
\Psi &=& \Psi_0 \exp[- \kappa t]
\label{exact_3.2}
.
\end{eqnarray}

Using the above system equations, 
the second-order numerical algorithm is given as follows. 
\begin{enumerate}
\item Advance the Stiff-part equations over $\Delta t / 4$ 
by using the formal solutions Eqs. (\ref{exact_1.2}) - (\ref{exact_3.2}).

\item Advance the non-stiff part of Maxwell equations 
Eqs. (\ref{4_Maxwell}) and (\ref{Maxwell_stiff})
over $\Delta t / 2$ by using method of characteristics as explained in Sec. \ref{sec:sec3.2}, 
and calculate numerical flux $F_{EM}$ (\ref{flux_EM_m}) and (\ref{flux_EM_E}). 
On the other hand, 
numerical flux $F_{fluid}$ (\ref{flux_fluid_D}) - (\ref{flux_fluid_E})
can be calculated by using approximate Riemann solver
~\citep{MM94,M96,B97,
ALO99,P00,F00,DB02,
MM03,MB05,M05}. 
In this paper, we use the HLLC solver. 

\item Advance conserved variables {D, ${\bf m}$, e} over the half time-step $\Delta t / 2$ 
by using Eqs. (\ref{RRMHD}) - (\ref{flux_EM_E}). 
Then, calculate primitive variables of half time step $U^{n+1/2}$ by primitive recovery 
explained in Sec. \ref{sec:sec3.5}. 
In our scheme, 
electric field $\mathbf{E}$ has to be evolved $\Delta t / 4$ by using formal solution 
(\ref{exact_1.2}) and (\ref{exact_2.2}) 
during primitive recovery. 
Primitive variables obtained through this procedure 
are used for the calculation of the numerical flux at $t = t + \Delta t / 2$. 

%\item Again, advance %stiff variables obtained through procedure 1 over $\Delta t / 4$
\item Again, advance initial stiff variables over $\Delta t / 2$
by using the formal solutions Eqs. (\ref{exact_1.2}) - (\ref{exact_3.2}).

%\item Calculate temporal second-order numerical flux 
\item Calculate temporal second-order numerical flux 
(\ref{flux_fluid_D}) - (\ref{flux_EM_E})
by using primitive variables obtained through the procedure 3. 
Then, advance conserved variables $D, \mathbf{m}, e$ over $\Delta t$
by Eq. (\ref{RRMHD}), and electric field $\mathbf{E}$ and magnetic field $\mathbf{B}$ 
by the Maxwell equations of stiff-part Eq. (\ref{Maxwell_stiff}). 

\item Calculate primitive variables by a primitive recovery process. 
During this process, 
the electric field $\mathbf{E}$ is advanced over $\Delta t / 2$ by using formal solutions 
(\ref{exact_1.2}) and (\ref{exact_2.2}).
\end{enumerate}

For the spatial second-order, 
we use the MUSCL scheme by Van Leer 
explained in Appendix.%Sec. \ref{sec:seca1}.

Note that 
if we evolve electric field ${\bf E}$ in integration of stiff equations 
or primitive recovery procedure, 
we have to evolve other primitive variables. 
This is because 
conserved variables are not changed during those procedures, 
and this means that 
the change of electric field ${\bf E}$ affects all other primitive variables. 
\section{Test simulations}\label{sec:sec4}
In this section, 
%several one-dimensional and multi-dimensional test simulations are presented. 
several one-dimensional test simulations 
given in previous studies K07 and P09 
are presented. 
For the numerical flux of fluid, 
we use the HLLC solver~\citep{MB05}. 
%We use an ideal equation of state with $\Gamma = 4 / 3$. 
We use an ideal equation of state 
$\rho \epsilon = p / (\Gamma - 1)$ 
with $\Gamma = 2$, 
and Courant number, $\mathrm{CFL} = 0.25$. 

%\subsection{\label{sec:sec4.1}One-dimensional test simulations}
%\subsubsection{\label{sec:sec4.1.1}Large amplitude CP Alfv\'en waves}
\subsection{Large amplitude CP Alfv\'en waves}\label{sec:sec4.1.1}
This test consists of the propagation of a large amplitude circularly-polarized Alfv\'en waves 
along a uniform background field $B_0$. 
The analytical exact solution of this problem is given by Del Zanna et al. (2007)
\citep{D07}; 
and this problem is used as the ideal-MHD limit test problem by P09. 
We use the same condition as P09. 
\begin{eqnarray}
(B^y, B^z) &=& \eta_A B_0 \left( \cos [k (x - v_A t) ], \sin[k (x - v_A t) ] \right)
,
\\
(v^y, v^z) &=& - \frac{v_A}{B_0} (B^y, B^z)
,
\end{eqnarray}
where $B^x = B_0$, 
$v^x = 0$, 
$k$ is the wave number, 
and $\eta_A $ is the amplitude of the wave. 
The special relativistic Alfv\'en speed $v_A$ is given by 
\begin{equation}
v_A^2 = \frac{2 B_0^2}{h + B_0^2 (1 + \eta_A^2)} 
\left( 1 + \sqrt{1 - \left( \frac{2 \eta_A B_0^2}{h + B_0^2 (1 + \eta_A^2)} \right)^2 } \right)^{-1}
. 
\end{equation}
For the initial data parameters, 
we have used $\rho = p = \eta_A = 1$, 
and $B_0 = 1.1547$.
Using these parameters, 
the Alfv\'en velocity is $v_A = 1 / 2$. 
For the boundary condition, 
the periodic one is used. 
In addition, 
we use a high uniform conductivity $\sigma = 10^6$ 
following P09, 
since this is the exact solution of ideal relativistic MHD. 

\begin{figure}[here]
%\plottwo{f2.eps}{f2_color.eps}
%\includegraphics[width=7cm]{CP_Alfven.eps} 
\epsscale{1.0}
\plotone{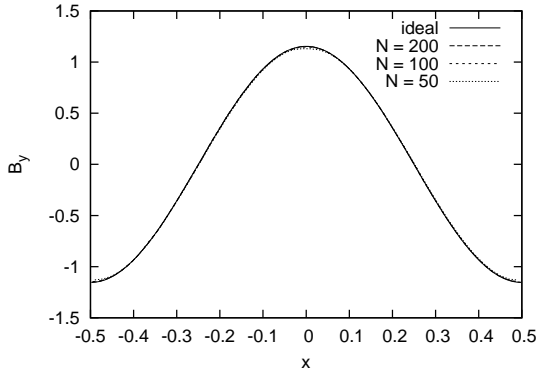} 
\caption{The results of large amplitude circularly-polarized Alfv\'en wave test 
         with large conductivity $\sigma = 10^6$. 
         This test is carried out for three different grid points $N = 50, 100, 200$.}
\label{fig1}
\end{figure}

Fig. \ref{fig1} is results of our new code at $t = 2.0$ (one Alfv\'en crossing time) 
for three different resolution cases with $N = \{50, 100, 200\}$. 
The computational domain is $x \in [-0.5, 0.5]$. 
This result indicates that 
our new code reproduces ideal relativistic MHD solutions 
when the conductivity $\sigma$ is high. 

\begin{figure}[here]
%\plottwo{f2.eps}{f2_color.eps}
%\includegraphics[width=7cm]{Convergence_Alfven.eps} 
%\includegraphics[width=7cm]{Convergence_current_sheet.eps} 
  \epsscale{1.0}
\plotone{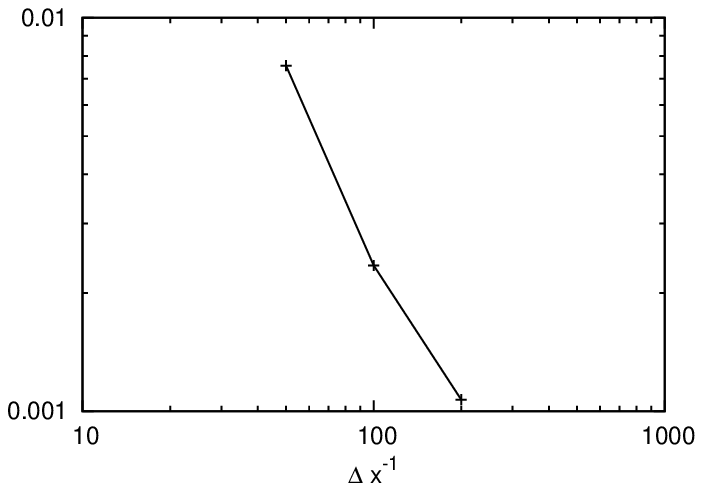}
\plotone{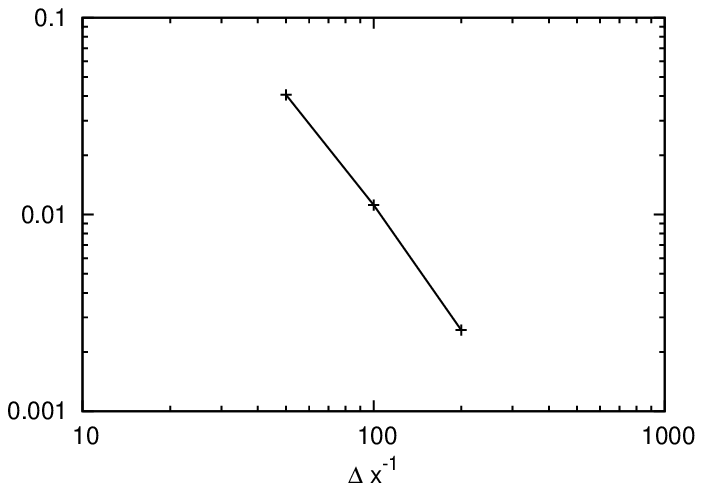} 
\caption{$L_1$ norm errors of the tangential magnetic field $B_y$ 
         under different grid resolution for the second-order schemes 
         using the new scheme. 
         The left-hand side is the result of Large amplitude CP Alfv\'en waves, 
         and the right-hand side is the result of the self-similar current sheet.}
\label{Convergence}
\end{figure}

%This test problem is smooth flow, 
%so that the convergence rate is consistent with the second-order spatial discretization. 
In these test problems, 
we cannot achieve full second-order accuracy. 
The left-hand side of Fig. \ref{Convergence} is the $L_1$ norm errors of 
the tangential magnetic field $B_y$ of this test problem. 
This figure shows that 
our numerical result is nearly $1.5$-order convergence. 
We estimate 
this is because our scheme uses many operator splittings, 
and the time accuracy of our scheme worsens. 
Note that 
this problem is one of the most difficult to solve in relativistic resistive MHD, 
since this is the limit of large conductivity $\sigma$. 
The right-hand side of Fig. \ref{Convergence} is the $L_1$ norm errors of 
the $B_y$ of the next test problem. 
Since the conductivity $\sigma$ is moderate value in that test problem, 
our new scheme achieves second-order convergence. 
%この問題でconvergence checkを行うと、
%時間２次精度は得られないが、１次以上の精度は得ることが出来た。
%これは私たちのコードはoperator splitを多用しているために、
%時間の２次精度を得ることが難しいことが原因であると考えられる。

%\subsubsection{\label{sec:sec4.1.2}Self-similar current sheet}
\subsection{Self-similar current sheet}\label{sec:sec4.1.2}
\begin{figure}[here]
%\plottwo{f2.eps}{f2_color.eps}
%\includegraphics[width=7cm]{current_sheet.eps} 
  \epsscale{1.0}
\plotone{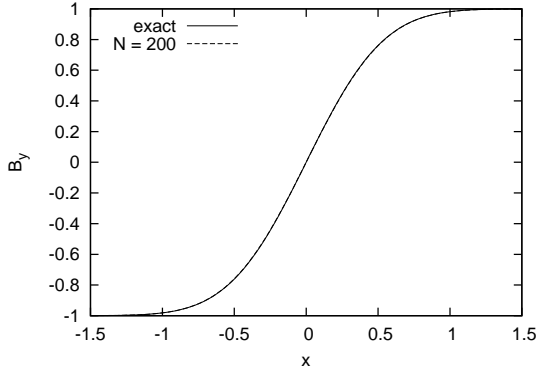} 
\caption{The result of self-similar current sheet test comparing the exact solution.
         The solid line is the exact solution, 
     and the dotted line is the numerical result with conductivity $\sigma = 10^2$.}
\label{fig2}
\end{figure}

This problem is used as the test problem of highly resistive cases 
in K07 and P09. 
%and detailed explanation of this test is given in them. 
In this test, 
it is assumed that 
the magnetic pressure is much smaller than gas pressure, 
so that the background fluid is not influenced by the evolution of the magnetic field. 
We assume 
the magnetic field has only tangential component $\mathbf{B} = (0, B(x,t), 0)$, 
and $B(x,t)$ changes its sign within this current sheet. 
Since we are interested only in the evolution of the magnetic field, 
the background fluid is set initially in equilibrium, $p = const$. 
In addition, 
we assume that the conductivity $\sigma$ is high, 
and the diffusion timescale is much longer than 
the light propagating timescale.
Although the resistive relativistic MHD equation is hyperbolic, 
this assumption allows us to neglect the displacement currents
at least in the rest frame. 
As the result, the evolution equation is reduced to 
\begin{equation}
\partial_t B - \frac{1}{\sigma} \partial_x^2 B = 0
.
\end{equation}
This equation has exact solution 
\begin{eqnarray}
B(x,t) &=& B_0 \mathrm{erf} \left( \frac{1}{2} \sqrt{\frac{\sigma}{\xi}} \right)
,
\\
\xi &=& \frac{t}{x^2}
, 
\end{eqnarray}
where erf is the error function. 
Following K07 and P09, 
we set the initial condition at $t = 1$ 
with $p = 50$, $\rho = 1$, $\mathbf{E} = \mathbf{v} = \mathbf{0}$, 
and $\sigma = 100$. 
The computational domain is $[-1.5, 1.5]$, 
and the number of grid points is $N = 200$. 
Fig. \ref{fig2} is the numerical result at $t = 9$. 
This figure shows that 
our scheme can solve a highly resistive problem accurately. 
The convergence rate is consistent with the second-order spatial and temporal discretization. 
%この図から分かるように、数値計算結果は厳密解と目視では差が分からないほど精度よく解けている。
%この問題でconvergence checkを行うと２次精度を得られていることが分かった。

%Since we use fast magnetosonic wave velocity as the characteristic velocity of the Maxwell equations, 
%the result is a little overshooted. 
%If we set speed of light as the characteristic velocity, 
%our code can reproduce the exact solution well 
%the same as Komissarov (2007) and Palenzuela et al. (2008). 

%\subsubsection{\label{sec:sec4.1.22}The propagation of Alfv\'enic transverse waves with Ohmic dissipation}
\subsection{The propagation of Alfv\'enic transverse waves with Ohmic dissipation}\label{sec:sec4.1.22}
%As explaind in Sec. \ref{sec:sec2}, 
%the relativistic resistive magnetohydrodynamic equation is 
%electromagnetic hydrodynamic equation with Ohmic dissipation. 
%This means that 
%our code has to reproduce both relativistic MHD in high conductivity regime 
%and 
%In this section, 
In order to confirm the capability of our method for the relativistic resistive MHD, 
we perform the test calculation of the propagation of Alfv\'enic transverse waves with Ohmic dissipation, 
%that becomes an Alfv\'en wave in the long wavelength regime, 
%and the light wave in the short wavelength regime, 
and compare the results with the exact dispersion relation Eq. (\ref{eq6}) 
%obtained in Appendix. \ref{sec:seca2}.
obtained in Appendix.% \ref{sec:seca2}.  if ApJ emulate

As explained in Appendix, 
the resistive relativistic magnetohydrodynamic equation contains transverse wave modes 
that become the light wave in large $k / \sigma$ region, 
and become the Alfv\'en wave in small $k / \sigma$ region. 
To demonstrate the propagation of transverse waves, 
we set the initial condition by eigenfunctions of the mode 
obtained from Eqs. (\ref{eqd1}) - (\ref{eqd4}) 
\begin{eqnarray}
  B^z &=& 0.05 \cos(k x), 
  \\
  v^z &=& \frac{B^x}{\rho h} \left\{ \left( \frac{\omega'}{k'} \right)^2 - 1 \right\} B^z, 
  \\
  E^x &=& \frac{B^x}{1 - i \omega'} v^z, 
  \\
  E^y &=& \frac{\omega'}{k'} B^z, 
\end{eqnarray}
where $\omega' \equiv \omega / \sigma$ and $k' \equiv k / \sigma = 2 \pi / \sigma$, 
and $\omega$ is the solution of the dispersion relation Eq. (\ref{eq6}). 
%For the small perturbation, 
%we set
%\begin{equation}
%  B_z = 0.05 \cos(k x)
%  ,
%\end{equation}
%where $k = 2 \pi$. 
%$v_z, E_x, E_y$ can be determined by using Eqs. (\ref{eqd1}) - (\ref{eqd4}), 
%and all the other fields are set to $0$. 
%We set the same parameters in Appendix \ref{sec:seca2}, 
We set the same parameters in Appendix.% \ref{sec:seca2}, if emulate ApJ
\begin{equation}
  (\rho h, B^x) = (1.5, 0.55)
  .
\end{equation}
Since the enthalpy includes the information of the equation of state, 
one can take any value of $\Gamma$. 
In this calculation, 
we set $\Gamma = 2$ and $p = 1$. 
The computational domain covers the region $[-0.5,0.5]$ 
where the periodic boundary condition is imposed, 
and the number of grid points is $N = 200$. 

%We calculate $1$ light crossing time when $k / \sigma > 1$, 
%and $1$ Alfv\'en crossing time when $k / \sigma \le 1$ by using ideal Alfv\'en velocity 
%Eq. (\ref{charact_Alfven_R}). 
The propagation speed of the numerical waves can be determined by tracing the position where $B_z$ is maximum. 
We measure the propagation speed and evaluate Re$[\omega]$ based on the time 
when the maximum of $B^z$ reaches $x = 0$ again, 
i.e. one-wave crossing period. 
%after $1$ Alfv\'en crossing time when $k / \sigma \le 1$ 
%and $1$ light crossing time when $k / \sigma > 1$. 
The damping rate Im$[\omega]$ is measured by using $B^z_M = B^z_0 \exp[\mathrm{Im}[\omega] t]$ 
where $B^z_M$ is the maximum of $B^z$ after the one-wave crossing time. 

\begin{figure}[here]
  \epsscale{1.0}
\plotone{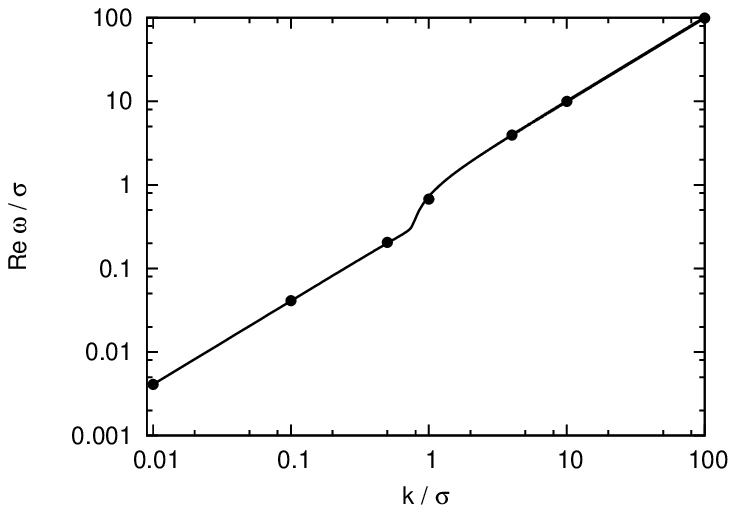}
\plotone{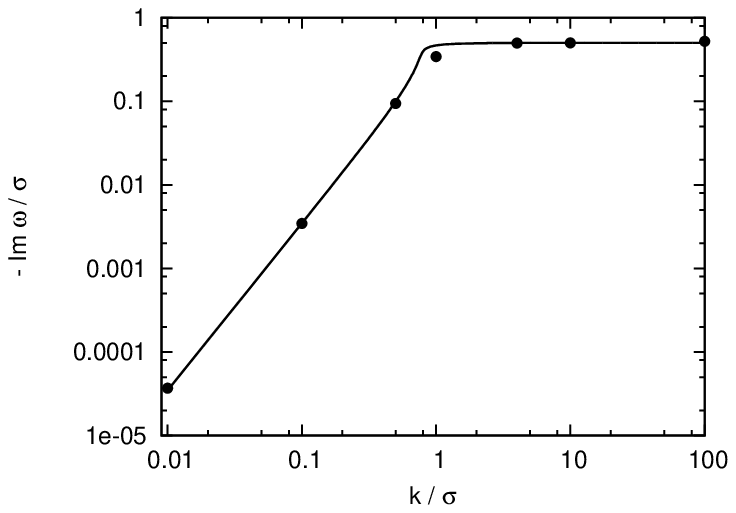} 
\caption{The result of the propagation of Alfv\'enic transverse waves with Ohmic dissipation test problem. 
         The solid line is the exact dispersion relation, 
         and the dots are the numerical solutions. 
         The test calculations are performed in the cases of 
         $k / \sigma = 0.01, 0.1, 0.5, 1, 4, 10, 100$.}
\label{fig_d}
\end{figure}

In Figs. \ref{fig_d}, 
we plot the real and imaginary part of $\omega / \sigma$ against $k / \sigma$. 
The solid line is the exact dispersion relation obtained in Appendix. 
We have performed the calculation in the cases of 
$k / \sigma = 0.01, 0.1, 0.5, 1, 4, 10, 100$. 
These figures show 
that our new numerical code can reproduce the propagation of Alfv\'enic transverse waves accurately 
for any value of the conductivity $\sigma$. 
%Since the damping time is nearly equal to propagation time in the case of $k / \sigma = 1$, 
%the 

%\subsubsection{\label{sec:sec4.1.3}Shock-tube problem}
\subsection{Shock-tube problem}\label{sec:sec4.1.3}
For the shock tube test problem, 
we consider the simple MHD version of the Brio and Wu test 
as P09. 
The initial left and right states are given by 
\begin{eqnarray}
(\rho^L, p^L, (B^y)^L) =& (1.0, 1.0, 0.5) & \mathrm{for} \quad x < 0.5 
\\
(\rho^R, p^R, (B^y)^R) =& (0.125, 0.1, -0.5) & \mathrm{for} \quad x \ge 0.5
\end{eqnarray}
All the other fields are set to $0$. 

\begin{figure}[here]
%\plottwo{f2.eps}{f2_color.eps}
%\includegraphics[width=7cm]{shock_tube_resolution.eps} 
  \epsscale{1.0}
\plotone{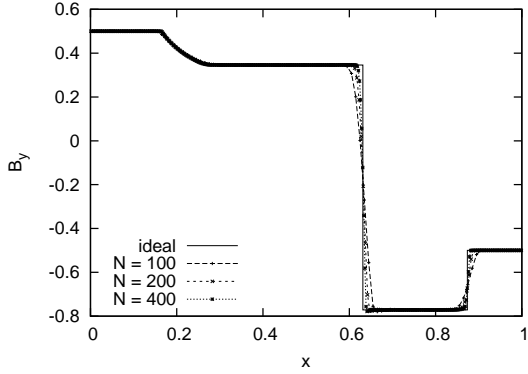} 
\caption{The numerical results of the Riemann shock tube test problem 
         for three different grid points $N = 100, 200, 400$.
         We use the conductivity $\sigma = 10^6$. 
         The solid line is the ideal solution.}
\label{fig3}
\end{figure}

Fig. \ref{fig3} is the numerical results at $t = 0.4$ 
that change grid points $N = 100, 200, 400$. 
The computational domain covers the region $[0, 1]$. 
We also plot an ideal RMHD solution by the solid line 
computed by a publicly available code developed by Giacomazzo and Rezzolla~\citep{GR06}. 
The conductivity is uniform with $\sigma = 10^6$. 
The solution of this Riemann problem contains a rarefaction moving to the left, 
a shock moving to the right, 
and a tangential discontinuity between them. 
Fig. \ref{fig3} shows that 
our numerical solution of the resistive MHD can reproduce the profile of an ideal MHD shock tube problem 
using high conductivity $\sigma$. 
In addition, 
our numerical solution captures contact discontinuity 
as sharp as P09. 
%まずFig. \ref{fig3}は$\sigma = 10^6$でメッシュ数を変えた場合の数値計算結果である。
%なおexact solutionはGiacomazzoからもらったコードを用いて計算している。
%この問題はrarefraction wave, shock, contact discontinuityが現れるが、
%いずれも精度よく計算出来ていることが分かり、
%contact discontinuityのcapture point数もKomissarov, Paleczuelaの計算結果と
%同等の結果が得られている。

\begin{figure}[here]
%\plottwo{f2.eps}{f2_color.eps}
%\includegraphics[width=7cm]{shock_tube_sigma.eps} 
  \epsscale{1.0}
\plotone{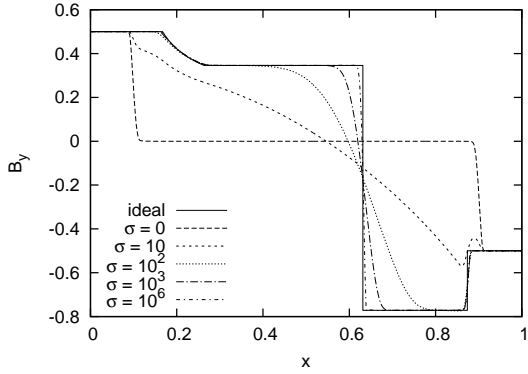} 
\caption{The numerical results of the Riemann shock tube test problem 
         for different conductivity cases: $\sigma = 0, 10, 10^2, 10^3, 10^6$. 
         The number of grid points is $N = 400$. 
         The solid line is the ideal solution.}
\label{fig4}
\end{figure}

Fig. \ref{fig4} is the numerical results of the same problem 
that changes the conductivity $\sigma = 0, 10, 10^2, 10^3, 10^6$. 
We also plot the ideal RMHD solution by the solid line. 
The number of grid points is $N = 400$. 
This result shows that 
our numerical solution reproduces nearly the same results as P09. 
%またFig. \ref{fig4}はメッシュ数を400に固定した場合に電気伝導度$\sigma$を変えた場合の
%数値計算結果である。Palenzuelaの計算結果と同等の結果が得られていることが分かる。

P09 reports that 
Strang's splitting method becomes unstable for moderately high values of the conductivity 
for this shock tube problem, 
and one has to use the implicit method. 
However, 
this is not related to whether one uses Strang's splitting or implicit method, 
but to the revision of the electric field during the iteration of the primitive recovery 
(H. R. Takahashi 2010, private communication). 
Our scheme uses Strang's splitting, 
but can solve this shock tube problem stably 
even when $\sigma \gtrsim 10^8$,  
if we revise the electric field during the primitive recovery 
as explained in Sec. \ref{sec:sec3.5}. 
%なおPalenzuelaはKomissarovのようにstiff partにexact solutionを用いた場合、
%解ける$\sigma$に上限があると報告しているが、これはexact solutionを用いるかimplicit法を
%用いるかには無関係であり、Primitive recoveryの際にPaleczuelaのように電場$E$を補正する
%ことが重要であることがわかった。
%私たちのコードはこれを組み込むことで$\sigma > 10^8$の場合にも解を得ることが出来る。

%Palenzuela says exact solution cannot solve high $\sigma$
%but modify as P we can solve $\sigma \ge 10^8$

\section{Test Simulations for fluid dominated case}\label{sec:sec5}
%The HLL code by Komissarov~\citep{Komissarov(2007)} uses velocity of light 
The previous studies K07 and P09 use light velocity 
for the characteristic velocity. 
Thus, 
%the accuracy of this code will become worse 
their numerical solutions become highly diffusive 
when one considers problems 
whose sound velocity or Alfv\'en velocity is much lower than light velocity. 
In this section, 
we perform test problems in such cases, 
and compare the results of the HLL code with that of our code. 
%KomissarovのHLL codeは光速を特性速度に用いているため、
%音速やAlfv\'en速度が光速に比べて小さいhigh densityの場合や
%high $\beta$の場合に精度が落ちることが
%予想される。
%この章ではこのような場合にテスト計算を行い結果を比較する。
\subsection{Shock tube test problem}\label{sec:sec5.1}
In this section, 
we compute a high plasma $\beta$ shock tube problem, 
and compare results of our code with those of the HLL code. 
%まずはhigh densityでhigh $\beta$の場合にshock tubeを計算する。
The initial left and right states are given by 
\begin{eqnarray}
(\rho^L, p^L, (B^y)^L) =& (10^4, 1.0, 0.05) & \mathrm{for} \quad x < 0.5 
,
\\
(\rho^R, p^R, (B^y)^R) =& (10^4, 0.1, -0.05) & \mathrm{for} \quad x \ge 0.5
.
\end{eqnarray}
All the other fields are set to $0$. 

\begin{figure}[here]
%\plottwo{f2.eps}{f2_color.eps}
%\includegraphics[width=7cm]{matter_dominated_shock_density.eps} 
%\includegraphics[width=7cm]{matter_dominated_shock_By.eps} 
  \epsscale{1.0}
\plotone{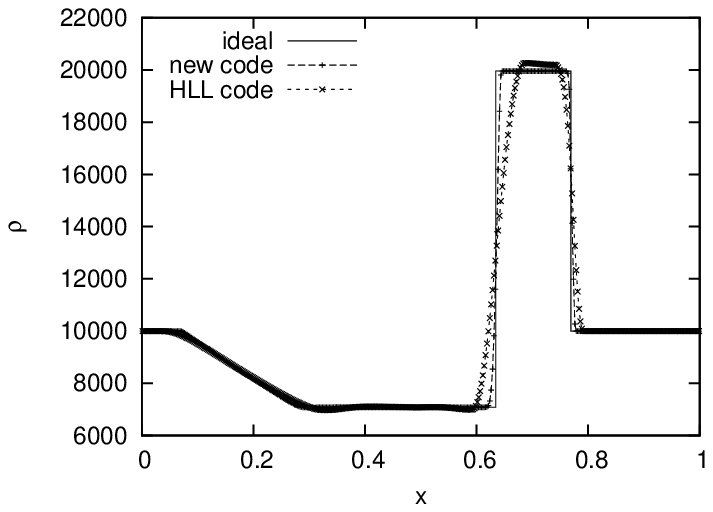}
\plotone{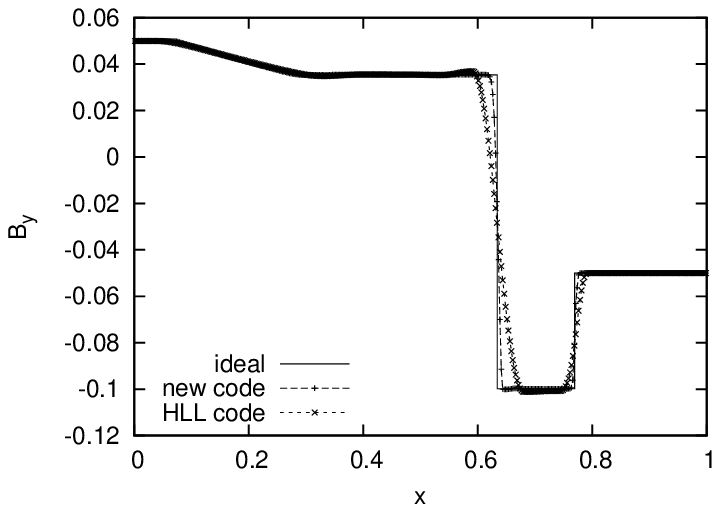} 
\caption{The numerical result of shock tube problem for fluid energy dominated case 
         comparing with that of HLL and ideal solution. 
         On the left is the density profile, 
         and on the right is the profile of the tangential magnetic field $B^y$. 
         The number of grid points is $N = 400$.}
%\caption{The tangential magnetic field $B^y$ profile of the numerical result of 
%         shock tube problem for fluid energy dominated case 
%         comparing with that of HLL and ideal solution. 
%         The number of grid points is $N = 400$.}
\label{fig5}
%\label{fig6}
\end{figure}

%Figs. \ref{fig5} and \ref{fig6} are the numerical results of our code and HLL one 
Figs. \ref{fig5} are the numerical results of our code and the HLL one, 
%of density and $B^y$ profiles 
being compared with ideal solutions at $t = 30.0$. 
The number of grid points is $N = 400$. 
%私たちのコードとHLLの結果をexact solutionと比較したのがFig. \ref{fig5}と
%Fig. \ref{fig6}で、それぞれ密度と$B^y$のprofileをプロットしている。
These figures show that 
HLL solver becomes more diffusive than our code. 
In addition, 
%Figs. \ref{fig5} and \ref{fig6} show that 
Figs. \ref{fig5} show that 
the density profile of the shock heated region somewhat overshoots that of the ideal solution, 
and tangential magnetic field $B^y$ slightly undershoots that of the ideal solution. 
These results show that 
when the plasma $\beta$ is high, 
the HLL solver becomes highly diffusive 
%the Komissarov's HLL solver becomes highly diffusive 
and does not reproduce the correct value of the shock heated region. 
In contrast, 
our numerical results reproduce ideal solutions very well 
even for high $\beta$ problems. 
%明らかにHLLの計算結果は精度が落ちていることが分かる。
%特にshock heated regionにおいて密度、$B^y$ともに間違った値を再現してしまっていることが分かる。
%それに対して私たちのコードは特性速度をきちんと扱っているために精度よく解けていることが分かる。

\subsection{The propagation of contact discontinuity}\label{sec:sec5.2}
In this section, 
we calculate the propagation of a contact discontinuity, 
and study the accuracy of capturing the contact discontinuity 
for various advection velocities. 
When one uses the HLL code by Komissarov, 
the numerical results can be expected to be diffusive 
for the case of very slow advection velocity, 
since the HLL code uses light velocity for the characteristic velocity. 
In contrast, 
our new code uses sound velocity for the fluid characteristic velocity, 
and the numerical results will be more accurate 
for any advection velocity. 
We consider the propagation of contact discontinuity of magnetohydrodynamics. 
The initial condition is 
\begin{eqnarray}
(\rho^L, p^L, (B^y)^L) =& (1.0, 1.0, 0.1) & \mathrm{for} \quad x < 0 
,
\\
(\rho^R, p^R, (B^y)^R) =& (1.5, 1.0, 0.05) & \mathrm{for} \quad x \ge 0
.
\end{eqnarray}
All the other fields are set to $0$. 

Since we want to consider the ideal fluid case, 
we consider high conductivity $\sigma = 10^6$. 
We use an equation of state with $\Gamma = 5 / 3$, 
and the computational domain covers the region $[-0.5,0.5]$ with $100$ grid points. 
The CFL number is $0.25$, 
and the integration is carried out until $2$ fluid crossing time. 
For the boundary condition, 
the periodic one is used. 
%計算は次の速度について行う。
For the advection velocity, 
we use the following velocities: 
\begin{equation}
  v^x = 0.9, 0.5, 0.1, 0.05, 0.01
  .
\end{equation}

%\begin{figure}[here]
%%\plottwo{f2.eps}{f2_color.eps}
%\includegraphics[width=7cm]{contact_B_new_v90_d.eps} 
%\includegraphics[width=7cm]{contact_B_new_v10_d.eps} 
%\includegraphics[width=7cm]{contact_B_new_v01_d.eps} 
%\includegraphics[width=7cm]{contact_B_new_v90_By.eps} 
%\includegraphics[width=7cm]{contact_B_new_v10_By.eps} 
%\includegraphics[width=7cm]{contact_B_new_v01_By.eps} 
%\caption{The numerical results of the propagation of contact discontinuity of RMHD 
%         for different advection velocity: $v^x = 0.9, 0.1, 0.01$. 
%         The number of grid points is $N = 100$.}
%  \label{fig9}
%\end{figure}
\begin{figure}[here]
  \epsscale{1.0}
\plotone{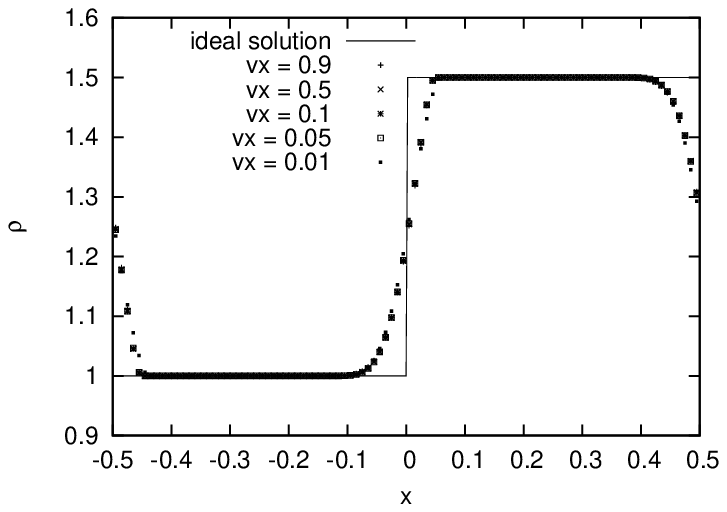}
\plotone{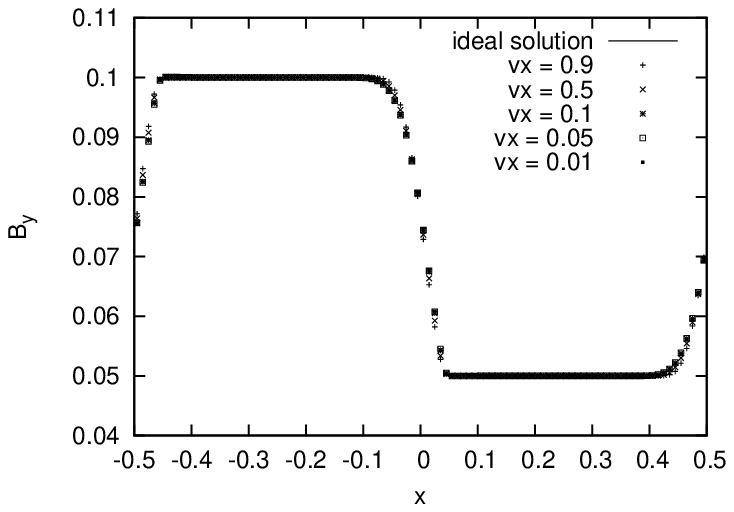} 
\caption{The numerical results of the propagation of contact discontinuity of RMHD 
         by using our new scheme 
         for different advection velocity: $v^x = 0.9, 0.5, 0.1, 0.05, 0.01$. 
         The number of grid points is $N = 100$.}
  \label{fig9_new}
\end{figure}
\begin{figure}[here]
\plotone{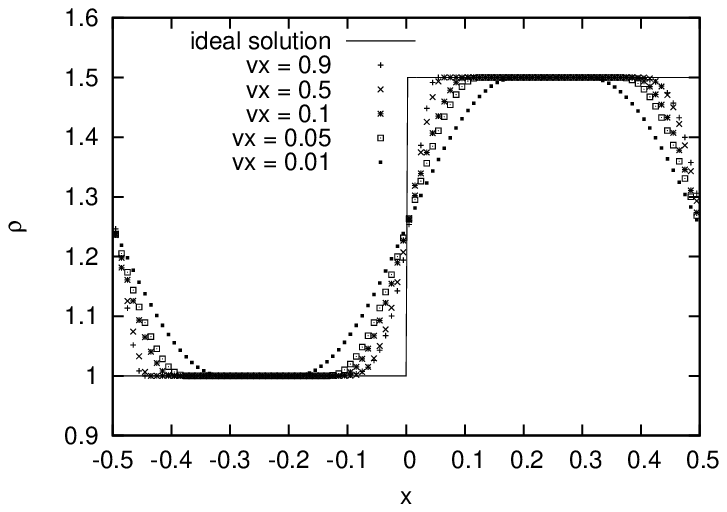}
\plotone{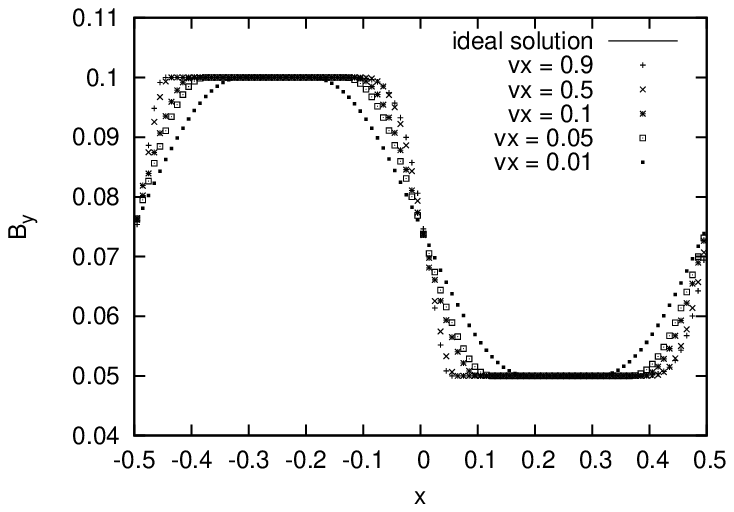} 
\caption{The numerical results of the propagation of contact discontinuity of RMHD 
         by using the HLL code 
         for different advection velocity: $v^x = 0.9, 0.5, 0.1, 0.05, 0.01$. 
         The number of grid points is $N = 100$.}
  \label{fig9_Kom}
\end{figure}
On the left of Figs. \ref{fig9_new} are the numerical results of the density profile 
calculated by using our new code, 
and on the left of Figs. \ref{fig9_Kom} are the numerical results of the density profile 
calculated by using the HLL code,  
The solid lines are the ideal solution. 
This figure shows that 
the numerical results of density profiles by HLL code of $v^x = 0.9$ 
is nearly equal to that of our new code. 
However, the numerical results by HLL code become more diffusive 
than by our code as the advection velocity becomes small; 
in contrast, 
the accuracy of the numerical results by our code is nearly independent of the advection velocity. 
The right hand side of Figs. \ref{fig9_new} are the numerical results 
of tangential magnetic field $B^y$ 
calculated by using our new code, 
and the right hand side of Figs. \ref{fig9_Kom} are the numerical results 
of tangential magnetic field $B^y$ 
calculated by using the HLL code.  
Similar to the density profile, 
the numerical results by using the HLL code become more diffusive 
than by using our code as the advection velocity becomes small; 
in contrast, 
the accuracy of the numerical results by our code is nearly independent of the advection velocity. 
%however, the numerical results by our code become more diffusive 
%as the advection velocity becomes small. 
%These results indicate that 
%the density profile of contact discontinuity by our code is independent of the advection velocity; 
%however, the tangential magnetic field of contact discontinuity by our code 
%strongly depends on the advection velocity, 
%and becomes more diffusive than Komissarov's HLL code 
%when the advection velocity is less than $0.1$. 
%This is because 
%we adopt the fast magnetosonic velocity as the characteristic velocity of electromagnetic field. 
%%The contact discontinuity does not contain shock, 
%%and the characteristic velocity is advection velocity, 
%In this problem, 
%the characteristic velocity is advection velocity. 
%however, 
%the fast magnetosonic wave is not equal to $v^x$ 
%even when the parallel magnetic field $B^x$ is $0$, 
%and the accuracy becomes bad 
%when the first term of the numerator of Eq. (\ref{charact_fm_EM}) is less than 
%the second term. 
%If one uses the Alfv\'en velocity as the characteristic velocity, 
%the numerical result of contact discontinuity becomes better; 
%however, 
%the accuracy of shock tube problem is a little worse than the fast magnetosonic velocity. 
%In addition, 
%the numerical result becomes even unstable 
%near the discontinuity. 
%これは恐らく特性速度としてfast magnetosonic waveを用いているため起こることで、
%移流の速度がAlfv\'en waveの速度を下回った場合、本来Bx=0ではないはずの情報の伝播が
%が$B^2$を用いているために発生しているためだと思われる。
%しかし衝撃波がある場合はfast magnetsonic waveの速度を用いた方が精度が良い。
%よって衝撃波があまり発生しないような亜音速の問題では我々のコードは精度が落ちるのでAlfv\'en waveを
%特性速度として用いるべきである。

In conclusion, 
the HLL code is not capable of accurately solving problems 
whose advection velocity is smaller than light velocity, 
since the HLL code uses light velocity for the characteristic velocity. 
%This result is not related to the contact discontinuity, 
%and Komissarov's HLL code becomes highly diffusive 
%when the characteristic velocity is much smaller than light velocity. 
The diffusive result of HLL code is always problematic for any discontinuity 
when the propagation velocity is much smaller than light velocity. 
In contrast, 
since our code uses appropriate characteristic velocities, 
the numerical dissipation does not depend on the characteristic velocity. 
For this reason, 
our new code can solve any advection velocity problems accurately, 
especially for the problems including discontinuities. 
%以上により、
%HLLコードは特性速度に光速を用いていることにより、
%移流速度が光速よりも遅い場合に数値散逸が極めて大きくなることが示された。
%これはcontact discontinuityの場合に限らずあらゆる現象に対しても同様のことが成り立ち、
%KomissarovのHLLコードは系の特性速度に光速よりも大きく遅いものがある場合に大きな数値拡散が現れ、
%構造をきちんと再現出来ないことになる。
%これに対し我々のコードは適した特性速度を用いているために、
%数値拡散は移流速度に関係がなくgrid数にのみ依存している。
%これにより、特に不連続面が存在する問題において我々のコードはHLLコードよりも極めて高い精度で
%計算出来ることになる。

\subsection{The propagation of small amplitude Alfv\'en wave}\label{sec:sec5.Slow}
In this section, 
we consider the propagation of small amplitude Alfv\'en waves 
in high $\beta$ plasma. 
The integration is performed for different resolutions, 
and we compare the numerical results of Komissarov's HLL code and our code. 
For the application to the numerical simulation of MRI, 
the integration is performed for a small number of grid points: $N = 16, 32, 64$ 
for one wavelength of the Alfv\'en wave; 
this corresponds to the number of grid points for resolving the wavelength of maximum growth rate of MRI. 
%この章では振幅の小さいAlfv\'en waveの伝播をhigh \beta$ plasmaについて解像度を変えて計算し、
%HLLスキームと私たちのコードの数値散逸率を比較する。
%解像度としてはMRIの最大成長波長を分解することをふまえて、
%$N = 16, 32, 64$の３通りを考える。

For the initial condition, 
we consider 
%初期条件としては次のものを考える。
\begin{eqnarray}
(\rho, p, B^x, B^y) &=& (10, 0.05, 0.1, 0.1) 
, 
\\
B^z &=& 0.01 \sin(2 \pi x / L)
,
\\
v^z &=& - \frac{B^z}{\sqrt{\rho h + |B|^2}}
.
\end{eqnarray}
In this case, 
the Alfv\'en velocity $v_A$ and plasma beta $\beta$ are given by 
\begin{equation}
  v_A = 3.14 \times 10^{-2}, \quad \beta = 5.02 \times 10^2
  , 
\end{equation}
where the Alfv\'en velocity and the plasma beta is defined as 
\begin{eqnarray}
v_A &=& \frac{B^x}{\sqrt{\rho h + |B|^2}}
, 
\\
\beta &=& \frac{\rho h}{|B|}
.
\end{eqnarray}
Since the initial magnetic field is very weak for most of the MRI phenomenon, 
a weak magnetic field is considered. 
%ここでMRIは初期に磁場が弱いため、初期条件としては弱い磁場を考えている。
%またAlfv\'en waveとしては次のような線形摂動の厳密解を考える。
%\begin{eqnarray}
%B^z &=& 0.01 \sin(2 \pi x / L)
%,
%\\
%v^z &=& - \frac{B^z}{\sqrt{\rho h + |B|^2}}
%,
%\end{eqnarray}
%相対論の場合、Alfv\'en速度 $v_A$とプラズマ $\beta$は次のものを用いる。
In order to consider the ideal fluid case, 
we set a high conductivity $\sigma = 10^6$. 
We use an equation of state with $\Gamma = 2$. 
The computational domain covers the region $[-0.5,0.5]$. 
The CFL number is $0.1$, 
and the integration is carried out until $1$ Alfv\'en wave crossing time. 
For the boundary condition, 
the periodic one is used. 
%The density, plasma $\beta$, and Alfv\'en velocity 
%are given by as follows: 
%%密度を変えた場合に減衰率について計算したものが次の表に示してある。
%\begin{center}
%  \begin{tabular}{ccc} \toprule
%   $\rho$    &   plasma $\beta$ & Alfv\'en velocity \\ \midrule
%   $0.1$ & 9.95 & $2.13 \times 10^{-1}$ \\
%   $1$ & 5.47 $\times 10^1$ & $9.45 \times 10^{-2}$ \\
%   $10$ & 5.02 $\times 10^2$ & $3.14 \times 10^{-2}$ \\
%%   $10^2$ & 4.98 $\times 10^3$ & $9.99 \times 10^{-3}$ \\
%%   $10^3$ & 4.98 $\times 10^4$ & $3.16 \times 10^{-3}$ \\
%%   $10^4$ & 4.98 $\times 10^5$ & $1.00 \times 10^{-4}$ \\ \bottomrule
%  \end{tabular}
%\end{center}
%%%%%%%%%%%%%%%%% 以下は表にする。\rho, \beta, v_A %%%%%%%%%%%%%%%%%%%%
%\begin{eqnarray}
%\rho &= 10^4, \qquad \beta &= 5 \times 10^5  
%,    
%\\   
%\rho &= 10^3, \qquad \beta &= 5 \times 10^4
%,    
%\\   
%\rho &= 10^2, \qquad \beta &= 5 \times 10^3  
%,    
%\\   
%\rho &= 10, \qquad \beta &= 5 \times 10^2
%,    
%\\   
%\rho &= 1, \qquad \beta &= 55
%,    
%\\   
%\rho &= 0.1, \qquad \beta &= 10
%,
%\\
%\end{eqnarray}
%%%%%%%%%%%%%%%%%%%%%%%%%%%%%%%%%%%%%%%%%%%%%%%%%%%%%%%%%%%%%%%%%%%%%
%For the diffusion rate, 
%For the damping rate, 
%we define the following damping of Alfv\'en wave per unit length 
%%減衰率はAlfv\'en waveの伝播を考えているため、
%%次のように距離辺りの減衰率を用いる。
%\begin{equation}
%  \kappa = \frac{|B^z_f| - |B^z_i|}{|B^z_i| c_A \Delta t} 
%  ,
%\end{equation}
%where $B_i$ is the initial value of magnetic field, 
%$B_f$ the final value of magnetic field, 
%and $c_A$ the Alfv\'en velocity. 
\begin{figure}[here]
  \epsscale{1.0}
\plotone{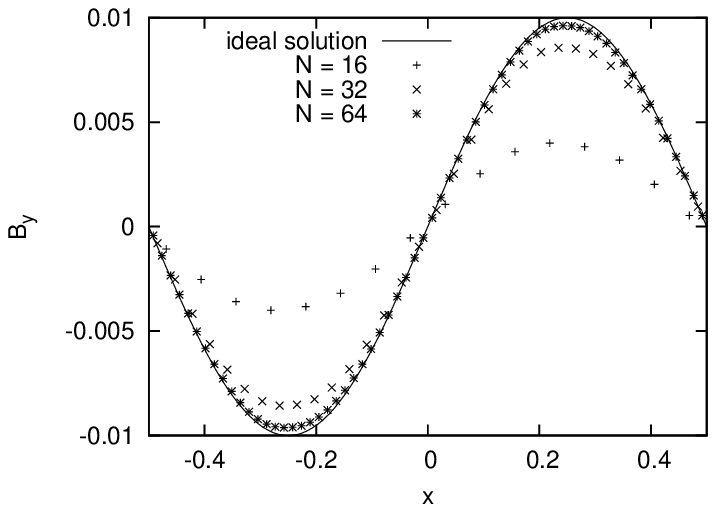}
\plotone{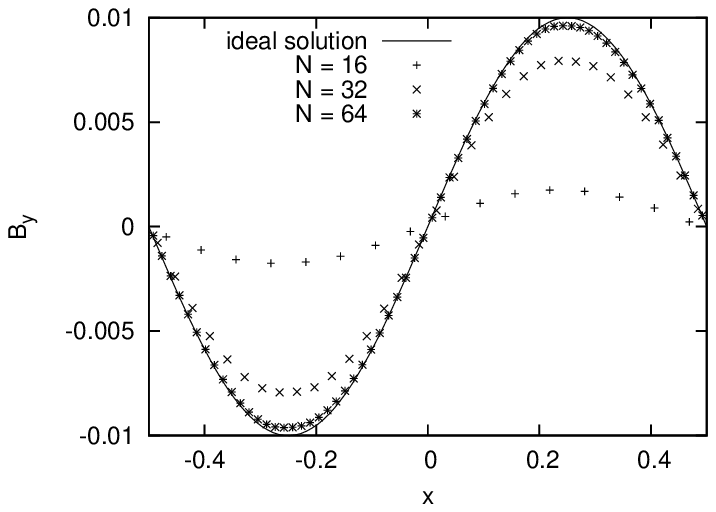} 
\caption{The numerical result of the propagation of a small slow Alfv\'en wave. 
%         in the case of $\rho = 10$.
         On the left is the result of our new scheme, 
         and on the right is that of the HLL code. 
         The number of grid points is $N = 16, 32, 64$. 
         }
\label{fig10}
\end{figure}

The numerical results of our code and HLL 
are presented in Figs. \ref{fig10}. 
Although the amplitude of both results falls 
because of the numerical diffusion, 
it can be seen that 
HLL results are more diffusive than our numerical results 
when the number of grid points is $N = 16, 32$. 
When the number of grid points is $N = 64$, 
the numerical result of HLL code is a little more accurate than that of our code. 
This is because 
our new scheme uses an operator split for the accuracy, 
and the convergence rate is a little less than second order in time. 
However, 
from a practical point of view, 
it is impossible to cost 64 grid points for the wavelength of maximum growth rate of MRI in many cases, 
and still our new code can integrate the growth of magnetic field by MRI more accurately. 

In conclusion, 
when one considers the high $\beta$ plasma, 
our code is more accurate than the HLL code 
because our code uses sound velocity and Alfv\'en velocity 
as the characteristic velocity. 
%以上により、high density, high $\beta$の場合は私たちのコードは
%特性速度を光速にして計算するHLLコードに比べて精度よく計算することが示された。
%The magnetorotational instability (MRI) is one of the most important astrophysical phenomena 
%that increases the magnetic fields exponentially. 
In particular, 
the above results show that 
our new method is useful for the application to the phenomena including MRI. 
%MRI is a magnetic dynamo phenomena in plasma. 
This instability occurs in the system 
whose angular momentum changes as $r^{-n} (0 < n < 2)$, 
and the amplitude of the perturbative Alfv\'en waves grows exponentially 
in over the duration of nearly one Kepler rotation. 
Since the above condition is satisfied in most of the differential rotating systems in gravity, 
MRI is one of the most important astrophysical phenomena. 
In order to reproduce this instability numerically, 
%one has to resolve the maximal grow wave length. 
one has to resolve the wavelength of maximum growth rate. 
However, 
this is difficult for most problems, 
since this wavelength is proportional to the initial weak magnetic field. 
For this reason, 
in order to reproduce MRI numerically, 
one has to use numerical schemes 
that can integrate small amplitude Alfv\'en waves accurately 
by smaller number of grid points. 
%宇宙流体において磁場を成長させる物理過程のうちで、
%MRIと呼ばれる不安定性がある。
%この不安定性は角速度が$r^{-n}$, $0 < n < 2$のように変化する系において
%Alfv\'en waveの振幅が１回転程度の時間で指数関数的に成長するという不安定性で、
%多くの重力系で発生するため非常に重要である。
%この不安定性を数値計算で再現するには、
%最大成長波長のAlfv\'en waveの伝播を解くことが必要であるが、
%この最大成長波長は初期磁場の大きさに依存し、
%多くの場合この波長を分解することは困難である。
%よってMRIをきちんと追うには、
%より少ない解像度で振幅の小さいAlfv\'en waveの伝播を解くことが出来るスキームを使うことが重要である。
Then, 
the results of test problems in this section show that 
our new numerical scheme can deal such problems more accurately than previous codes. 

In these three test problems, 
we consider extremely high density cases 
in order to distinguish differences easily. 
However, 
this can always happen 
when the magnetic field is weak. 
As a result, 
if one considers problems including an initially weak magnetic field 
like magnetic rotational instability (MRI) in the accretion disk, 
our code can produce more accurate results. 

%ここでは差を見やすくするために極端にhigh densityの場合を考えた。
%しかし磁場が小さい場合は常にこのような状況は起きうるため、
%MRIのように初期にhigh $\beta$であるようなplasmaの問題では私たちのコードの方が良い精度で
%解けることがわかる。

%\subsection{\label{sec:sec4.2}Multi-dimensional test simulations}

\section{Conclusion}\label{sec:sec6}
In this paper, 
we have presented a new numerical scheme of resistive RMHD for one-dimensional case 
which can solve matter dominated problems more accurately 
than the existing numerical method. 
Since this new scheme uses different characteristic velocity 
for obtaining the numerical flux of fluid and electromagnetic field, 
one can solve accurately and stably problems 
whose characteristic velocity is much lower than that of light. 
%This new scheme splits the numerical flux 
%into fluid part and electromagnetic part; 
%for fluid numerical flux, 
%we use Riemann solver
%and 
%for electromagnetic numerical flux, 
%we use method of characteristics. 
%この論文では新しい相対論的散逸磁気流体のスキームが与えられた。
%このスキームでは数値流速を求めるために流体部分と電磁場を分離し、流体部分は特性速度
%として音速を用いた近似Riemann solverを用い、
%電磁場は特性速度としてfast magnetosonic velocityを用いた特性曲線法で計算する。
%この結果HLLコードとは違い、high density, high $\beta$のように特性速度が大きく
%光速から異なる問題も精度よく安定に解くことが出来る。

When one considers relativistic problems, 
one has to solve stiff equations for electric fields. 
In general, 
it is difficult to deal with stiff equations, 
and special methods have been presented; 
for example, 
K07 uses the Strang's splitting method, 
and P09 use the implicit method. 
P09 report that 
Strang's splitting method is incapable of solving problems 
that include discontinuity, such as shock. 
However, 
we find that 
this is not related to the method for the stiff equations, 
and one can solve problems including shock 
if one evolves the electric field during the primitive recovery; 
we use Strang's splitting method, 
and the solver is well behaved for shock tube problems. 
The results of other test problems show that 
our new scheme is capable of accurately solving 
both highly resistive problems and nearly ideal MHD ones. 
In addition, 
it has been shown that 
our code can solve low characteristic velocity problems more accurately 
than the HLL code. 
%our code can solve accurately low characteristic velocity problems 
%that cannot be solved by the HLL code presented by Komissarov. 
%相対論的な場合、resistive MHDは非相対論とは違い因果律を守るために電場も
%考慮して解く必要がある。
%この場合ideal MHD近似が成り立つ場合、電場の時間発展がstiffになり、
%精度よく解くのは大変難しい。
%この問題に対して、KomissarovがStrang's splitting methodを、
%Palenzuela el alがimplicit methodをそれぞれ用いて解決している。
%一方Palenzueltaはstrang's splitting methodを用いた場合は、
%ショックのような強い衝撃波がある場合は不安定になり解けないと報告していたが、
%私たちはこの原因はスキームの選択ではなく、primitive recoveryの際に
%電場の時間発展を考慮するかが問題であることを発見した。
%私たちのコードはStrang's splitting methodを用いているが、
%primitive recoveryの際に電場の時間発展を考慮することで
%衝撃波が存在する問題も精度よく解くことが出来ている。
%その他resistivityが強い場合やMHD limitなど様々なテスト計算をしたが、
%精度よく数値解が得られている。
%更にHLLコードでは解けないような、特性速度が光速より極端に小さいような問題も
%精度よく解けることが示された。

The problems of high density and high plasma $\beta$ appear 
when one considers magnetorotational instability (MRI) 
  in the accretion disk with a relativistic jet, for example. 
In this case, 
one has to use relativistic resistive MHD code 
that can solve both highly relativistic and non-relativistic dynamics with resistivity 
for the following three reasons: 
(1) the saturation of the magnetorotational instability (MRI) depends on the resistivity; 
(2) the dynamics of an accretion disk are not ordinarily relativistic, 
especially, 
the dynamics of the MRI is sub-Alfv\'enic; 
(3) the dynamics of the jet are highly relativistic. 
Our new scheme can solve such problems accurately 
even when the initial magnetic field is very weak. 
%We will solve these problems in our next work. 

%In our next paper, 
%we present multi-dimensional extension of our scheme. %, 
%and present some astrophysical application. 
We present multi-dimensional extension of our scheme in our next paper. %, 
%When one solves ideal MHD problems in a differentially rotating system,  
%MRI inevitably happens 
%and any little initial magnetic field grows exponentially. 
%Accretion disks around black holes are a hot perfectly ionized plasma, 
%and are expected that it can be approximated ideal MHD and is unstable for MRI. 
%In particular, 
%when one solves the dynamics of the accretion disk around black hole 
%with a relativistic jet, 
%one has to use a relativistic resistive MHD code 
%that is capable of solving both highly relativistic and non-relativistic dynamics. 
%
%In addition, 
%we use scalar Ohm's law in this paper, 
%and it cannot be used for strong plasma. 
%Thus, 
%私たちのコードが精度よく解くことの出来るhigh density, high $\beta$な状況は
%例えばaccretion diskのMRIを考える場合に重要になる。
%MRIはaccretion diskのような微分回転をする系において、
%理想MHDの場合、初期磁場がどんなに小さくても流体とequipartition程度まで磁場が成長するという
%特徴を持つ不安定性である。
%BH周りのaccretion diskの場合は、通常完全電離プラズマなのでresistivityは非常に小さく、
%MRIにより磁場は必ず成長すると期待される。
%数値計結果からも分かるようにHLLコードはこのような場合に精度よく計算が出来ないが、
%私たちのコードはこのような場合でも流体変数、電磁場の両方を精度よく計算が出来る。
%私たちは次の論文でこのような場合について計算する予定である。
%また今回採用したオームの法則は最も単純なもので、磁場が極端に強く非等方が強い場合には
%使えないため、今後はこの問題も解決する必要がある。

%% Included in this acknowledgments section are examples of the
%% AASTeX hypertext markup commands. Use \url without the optional [HREF]
%% argument when you want to print the url directly in the text. Otherwise,
%% use either \url or \anchor, with the HREF as the first argument and the
%% text to be printed in the second.

\acknowledgments
We would like to thank Bruno Giacomazzo for providing the code computing 
the exact solution of the Riemann problem in ideal MHD. 
Numerical computations were in part carried out on Cray XT4 
at Center for Computational Astrophysics, CfCA, of National Astronomical Observatory of Japan.
This work is supported by Grant-in-aids from the Ministry of Education, Culture, Sports, Science, 
and Technology (MEXT) of Japan, No. 22$\cdot$3369 (T. I.). 


%% in parentheses, after the keyword, but they will not be verified.

%{\it Facilities:} \facility{Nickel}, \facility{HST (STIS)}, \facility{CXO (ASIS)}.



\appendix

%\section{Appendix}
%\subsection{\label{sec:seca2}The dispersion relation of the relativistic electromagnetic fluid}
\section{The dispersion relation of the relativistic electromagnetic fluid}\label{sec:seca2}
As explained in Sec. \ref{sec:sec3.2}, 
we solve the evolution of electric and magnetic field 
by the method of characteristics. 
For the characteristic velocity, 
we use the appropriate MHD characteristic velocity 
%we use the fast magnetosonic velocity 
%we use the Alfv\'en velocity 
when $\sigma$ is large, 
that is, ideal MHD approximation is valid. 
%when the MHD is very good approximation. 
However, 
when the conductivity $\sigma$ is not so large, 
we have to replace the characteristic velocity with the speed of light. 
In this section, 
we discuss 
when to switch the characteristic velocity from appropriate MHD characteristic velocity to light velocity. 
In the following, 
we calculate the linear perturbation of the relativistic electromagnetic equation 
in order to obtain characteristic velocity. 
%when to switch the characteristic velocity from fast magnetosonic velocity to light velocity. 
%when to switch the characteristic velocity from Alfv\'en velocity to light velocity. 
%私たちのアルゴリズムは電磁場を特性曲線法によって解くが、
%特性速度としてfast magnetosonic velocityを用いている。
%$\sigma \gg 1$のideal MHD近似が極めて良い近似になる場合はMHDの場合の
%fast magnetosonic velocityを用いればいいが、resistivityが小さい場合は
%MHD近似が悪くなるため、特性速度として光速を使わなければならない。
%この章ではどのような場合に特性速度を光速に切り替えるのかについて議論する。

The relativistic electromagnetic fluid equations are given by 
%まずREMHD方程式は次で与えられる。
\begin{eqnarray}
\rho h u^{\mu} \partial_{\mu} \mathbf{u}^i &=& - \nabla p 
+ (q \mathbf{E} + \mathbf{J \times B})
\label{eq1}
,
\\
\partial_t \mathbf{B} &=& - \nabla \times \mathbf{E}
\label{eq2}
,
\\
\partial_t \mathbf{E} &=& \nabla \times \mathbf{B} - \mathbf{J}
\label{eq3}
,
\\
\mathbf{J} &=& \sigma \gamma [\mathbf{E} + \mathbf{v \times B} - (\mathbf{E \cdot v}) \mathbf{v}] 
+ q \mathbf{v}
\label{eq4}
,
\\
q &=& \nabla \cdot \mathbf{E}, \qquad \nabla \cdot \mathbf{B} = 0
\label{eq5}
.
\end{eqnarray}

To obtain the dispersion relation, 
we start by expanding physical variables 
around an unperturbed state 
in the following frame: 
%この式を線形摂動する。
%この際一般に次のように仮定出来る。
\begin{itemize}
\item
The fluid is at rest: 
%流体は静止している : 
$\mathbf{v}_0 = \mathbf{0}$
\item
The x-coordinate is parallel to the ${\bf k}$:
%x軸は波数ベクトル方向 : 
$\mathbf{k} = k \mathbf{e}_x$
\item
The magnetic field is in the x-direction:  
%磁場はx-y平面内 : 
$\mathbf{B}_0 = B^x \mathbf{e}_x$
\item
charge neutrality: 
%電荷中性 : 
$q_0 = 0, \quad \mathbf{E}_0 = {\bf 0}$
\end{itemize}
Since we only want to judge 
when to switch characteristic velocity, 
we consider propagation of the transverse waves along the magnetic field. 
When one uses this procedure during the numerical simulation, 
one only has to calculate $B^2 - E^2$ of the simulation data, 
and substitute its square root into the above $B^x$. 
This is because $B^2 - E^2$ is a scalar, 
and becomes the square of the magnetic field in the fluid comoving frame 
because of the assumption of the charge neutrality. 
Since the magnetic field appears only in the form of $B^2$ in the following procedure, 
one can neglect the sign of magnetic field. 
In the following, 
we consider only the characteristic velocity of transverse waves. 
%Since we only want to judge when to switch characteristic velocity, 
%we set magnetic field in the x-direction. 
%In the following, 
%we consider only the characteristic velocity of Alfv\'en mode. 

In the above condition, 
the Alfv\'en mode is included in the z-component of the velocity $\delta v^z$ and magnetic field $\delta B^z$, 
and decouples from other variables. 
For this reason, 
we consider only variables related to $\delta v^z$ and $\delta B^z$. 
%このような条件の場合、Alfv\'en速度は速度$\mathbf{v}$と磁場$\mathbf{B}$のz成分に
%現れるため、これらに関係する量のみ線形摂動を考える。

We replace the current vector in Eq. (\ref{eq1}) with Eq. (\ref{eq3}). 
Then, the perturbed equations are 
%まず電荷中性よりEq. (\ref{eq1})の電場の寄与は線形には寄与しない。
%次にEq. (\ref{eq1})の電流にEq. (\ref{eq3})を代入する。
%すると次のような式が得られる。
\begin{eqnarray}
i \omega \rho h \delta v^z + i k B^x \delta B^z - i \omega B^x \delta E^y = 0
%i \omega \rho h \delta v^z + i k B^x \delta B^z + i \omega B^y \delta E^x - i \omega B^x \delta E^y = 0
\label{eqd1}
,
\\
i \omega \delta B^z - i k \delta E^y = 0
\label{eqd2}
,
\\
\sigma B^x \delta v^z + (i \omega - \sigma) \delta E^x = 0
\label{eqd3}
,
\\
\sigma B^x \delta v^z + i k \delta B^z + (\sigma - i \omega) \delta E^y = 0
\label{eqd4}
.
\end{eqnarray}
From these equations, 
%If the determinant of the above homogeneous system is set to equal zero, 
the following dispersion relation is obtained: 
%この方程式が自明な解以外を持つ条件から次の分散関係式が求まる。
\begin{equation}
\rho h \omega^4
+ i \sigma (B^2 + 2 \rho h) \omega^3 
- [k^2 \rho h + \sigma^2 (B^2 + \rho h) ] \omega^2 
- i \sigma (B^2 + \rho h) k^2 \omega 
+  \sigma^2 (B^x)^2 k^2 = 0
,
\label{eq6}
\end{equation}

\begin{figure}[here]
%\plottwo{f2.eps}{f2_color.eps}
%\includegraphics[width=7cm]{charact_r_3.eps} 
%\includegraphics[width=7cm]{charact_i.eps} 
  \epsscale{1.0}
\plottwo{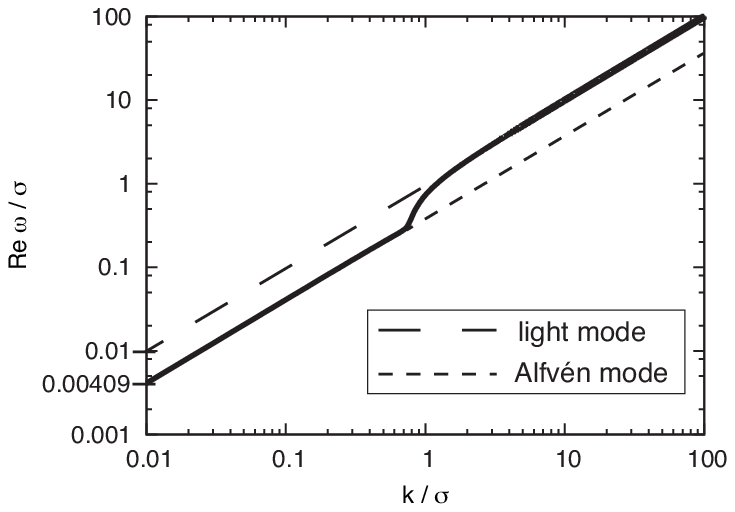}{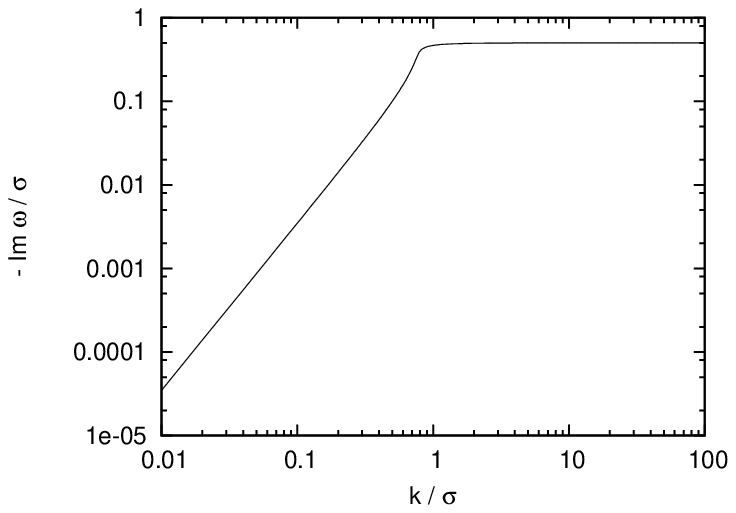} 
\caption{The dispersion relation of the propagation mode. 
         The left-hand side is the real part of $\omega / \sigma$, 
         and the right-hand side is the imaginary part of $\omega / \sigma$. 
         In the figure of the real part of $\omega / \sigma$, 
         we also plot reference lines 
         whose phase velocities are Alfv\'en velocity (long-dashed line) 
         and speed of light (short-dashed line). 
         The phase velocities can be obtained from the data at $k / \sigma = 0.01$ 
         by using the formula $c_{phase} = \omega / k$. 
         The parameters are set as $(\rho h, B^x) = (1.5, 0.55)$, 
         and the Alfv\'en velocity is given by $v_A \simeq 0.409$. 
         This figure shows that 
         this mode becomes light waves in the limit of large $k / \sigma$ 
         and Alfv\'en waves in the limit of small $k / \sigma$. 
         In addition, 
         this mode has a maximum decay rate in the limit of large $k / \sigma$.}
%\caption{The imaginary part of the dispersion relation of the propagation mode. 
%         This figure shows that 
%         this mode have maximum decay rate in the limit of large $k / \sigma$.}
\label{figa1}
%\label{figa2}
\end{figure}

Eq. (\ref{eq6}) is the biquadratic equation with respect to $\omega$, 
and has the formula of radicals. 
However, 
%this formula needs a solution of a cubic equation, 
the analytical formula is very complex and hard to analyze, 
%In addition, 
%this formula is numerically unstable for some parameter of our problem. 
and is not suitable for obtaining the characteristic velocity. 
%Thus, that is not suitable for obtaining the characteristic velocity. 
%Instead of solving above biquadratic equation directly, 
%we determine the characteristic velocity 
%that decays after $\Delta t^n$ obtained by the CFL condition. 
%If it is light velocity, 
%we use it as the characteristic velocity; 
%otherwise, 
%we use fast magnetosonic velocity as the characteristic velocity. 
%we use Alfv\'en velocity as the characteristic velocity. 
As shown below, 
the transverse wave becomes an Alfv\'en wave in the long wavelength regime, 
%the Alfv\'en mode becomes Alfv\'en wave in the long wavelength regime, 
and the light wave in the short wavelength regime 
shown in Fig. \ref{figa1}. 
Fig. \ref{figa1} shows that 
the damping rate is a monotonically increasing function of $k$. 
For this reason, 
we establish the following criterion for the characteristic velocity: 
when all light modes damp during one time step $\Delta t^n$ 
we use appropriate MHD characteristic velocity for the method of characteristics; 
when some light modes do not damp during one time step $\Delta t^n$, 
we use light velocity for the method of characteristics. 
We will discuss this method in detail in the following.
%この式は$\omega$に対する4次方程式なので解の公式が存在するが、
%その解は中に３次方程式の解を含むなど解析的にかなり複雑な形をしていて、
%さらにパラメータによっては数値的に不安定になるため
%直接特性速度を求めるには向かない。
%そこで私たちのアルゴリズムでは、直接特性速度を求めることはせずに、
%Courant conditionから求まった$\Delta t$で減衰する場合の特性速度を
%求め、それが光速の場合に、特性曲線法の特性速度を光速に切り替えることにする。

First, 
we substitute $\omega = \omega_R + i \omega_I$ into Eq. (\ref{eq6}), 
and divide the dispersion relation into a real part and imaginary part. 
Then, 
the real part is 
%そのためにまずEq. (\ref{eq6})に$\omega = \omega_R + i \omega_I$を
%代入し、分散関係式を実部と虚部に分ける。
%まず実部は
\begin{eqnarray}
&&\rho h \omega_R^4
- [\rho h k^2 + \sigma^2 (B^2 + \rho h) 
+ 3 \sigma (B^2 + 2 \rho h) \omega_I + 6 \rho h \omega_I^2] \omega_R^2 
\nonumber
\\
&+& \rho h \omega_I^4 
+ \sigma (B^2 + 2 \rho h) \omega_I^3 
\nonumber
\\
&+& [\rho h k^2 + \sigma^2 (B^2 + \rho h) ] \omega_I^2 
+ \sigma (B^2 + \rho h) k^2 \omega_I 
+ (B^x)^2 \sigma^2 k^2 = 0
,
\label{eq7}
\end{eqnarray}
and the imaginary part is 
%虚部は
\begin{equation}
[B^2 \sigma + 2 \rho (\sigma + 2 \omega_I) ] \omega_R^3
- [\rho h (\sigma + 2 \omega_I) \{k^2 + 2 \omega_I (\sigma + \omega_I) \} 
+ B^2 \sigma \{k^2 + \omega_I (2 \sigma + 3 \omega_I) \} ] \omega_R =0
,
\label{eq8}
\end{equation}
Eq. (\ref{eq8}) implies that 
the solution for $\omega_R$ is $0$ and a conjugate complex numbers. 
%and this is checked by the numerical experiments 
%for some parameters. 
The solutions of $\omega_R = 0$ are pure decaying modes, 
so the other modes are the desired propagating ones 
that become light velocity in the limit of small $\sigma$ 
and Alfv\'en velocity in the limit of large $\sigma$. 
%虚部の方程式Eq. (\ref{eq8})より、$\omega$の実部はある複素数とその共役量、$0$で与えられる
%ことがわかる。これは数値的実験によっても確かめられた。
%$0$は純粋な減衰解であるため、REMHDでは伝播モードは$\sigma$が小さい場合は光速に、$\sigma$が大きい
%場合にはAlfv\'en速度になる減衰波であることが分かる。
Figs. \ref{figa1} are the dispersion relation for the propagation modes 
of the following parameters: 
%Fig. \ref{figa1}, \ref{figa2}は次のパラメータの場合の伝播モードの分散関係である。
\begin{equation}
  (\rho h, B^x) = (1.5, 0.55)
%  (\rho, h, B^x) = (1.5, 1.0, 0.55)
%  (\rho, h, B^x, B^y) = (0.5, 1.0, 0.5, 0.1)
\end{equation}
These figures show that 
this mode becomes light in the limit of small $\sigma$ 
and Alfv\'en wave in the limit of large $\sigma$. 
Although the form of $\omega_R$ does not become as Fig. \ref{figa1} for some parameters, 
this mode always becomes light waves 
in the limit of small $\sigma$. 
%このように長波長、もしくは$\sigma$が大きい場合はAlfv\'en waveに、
%短波長、または$\sigma$が小さい場合は光速に帰着することが分かる。
%パラメータによっては必ずしもこのようなグラフの形にはならないが、
%いずれのパラメータでも短波長、$\sigma$が小さい場合では必ず光速に帰着する。

From Eq. (\ref{eq8}), 
this desired mode can be obtained as follows: 
%私たちが欲しいのはこの伝播モードであるので、Eq. (\ref{eq8})から次の$\omega_R^2$である。
\begin{equation}
\omega_R^2 = \frac{\rho h (\sigma + 2 \omega_I) [k^2 + 2 \omega_I (\sigma + \omega_I)] 
+ B^2 \sigma [k^2 + \omega_I (2 \sigma + 3 \omega_I)]}
{B^2 \sigma + 2 \rho h (\sigma + 2 \omega_I)}
.
\label{eq9}
\end{equation}
We substitute this $\omega_R^2$ into Eq. (\ref{eq7}), 
%and it reduces to 
and we obtain 
%この$\omega_R^2$を実部の式Eq. (\ref{eq7})に代入し、整理する。
\begin{equation}
\alpha_4 k^4 + \alpha_2 k^2 + \alpha_0 = 0
,
\label{eq10}
\end{equation}
where
\begin{eqnarray}
\alpha_4 &=& - B^2 \rho^2 h^2 \sigma (\sigma + 2 \omega_I) - \rho^3 h^3 (\sigma + 2 \omega_I)^2
,
\\
\alpha_2 &=& B^4 (B^x)^2 \sigma^4 - B^6 \sigma^3 (\sigma + 2 \omega_I) 
- 2 B^4 \rho h \sigma^2 (2 \sigma^2 + 7 \sigma \omega_I + 6 \omega_I^2) 
\nonumber
\\
&-& \rho^2 h^2 (\sigma + 2 \omega_I)^2 \{- 4 (B^x)^2 \sigma^2 + 2 \rho h (\sigma + 2 \omega_I)^2\}
\nonumber
\\
&-& B^2 \rho h \sigma (\sigma + 2 \omega_I) 
\{- 4 (B^x)^2 \sigma^2 + \rho h (5 \sigma^2 + 18 \sigma \omega_I + 16 \omega_I^2)\}
,
\\
\alpha_0 &=& - 2 B^6 \sigma^3 \omega_I (\sigma + 2 \omega_I)^2 
- 4 \rho^3 h^3 \omega_I (\sigma + \omega_I) (\sigma + 2 \omega_I)^4 
\nonumber
\\
&-& 2 B^2 \rho^2 h^2 \sigma \omega_I (\sigma + 2 \omega_I)^3 (5 \sigma + 6 \omega_I) 
\nonumber
\\
&-& 4 B^4 \rho h \sigma^2 \omega_I (\sigma + 2 \omega_I) (2 \sigma^2 + 7 \sigma \omega_I + 6 \omega_I^2) 
.
\end{eqnarray}
This equation should include the propagation modes. 

Eq. (\ref{eq10}) is the biquadratic equation with respect to $k$, 
but includes unknown quantity $\omega_I$. 
%Then, it should be noticed that 
%what we want to obtain is the characteristic velocity 
%whose decay time is less that $\Delta t^n$ 
%determined by the CFL condition. 
Figs. \ref{figa1} show that 
the propagation mode becomes light waves in the short wavelength region, 
and damping rate $- \omega_I$ is a monotonically increasing function of $k$. 
Note that 
what we want to know is 
whether the undamped shortest wavelength mode is light waves or Alfv\'en waves, 
and we do not necessarily have to solve the biquadratic equation directly. 
%Thus, 
%if one obtains the wave number $k$ 
%whose decay time is equal to $\Delta t^n$, 
%the maximum velocity of the wave 
%that can propagate without decaying 
%can be obtained 
%by using this wave number $k$ and $\omega_R(k)$. 
%この式は伝播モードのみ含んでいるはずである。
%この式は$k$の４次方程式になっているが、未知数$\omega_I$を含んでいる。
%ここで私たちが欲しいのは、Courant conditionで決まる$\Delta t$より減衰時間が小さいモードの特性速度である。
%Fig. \ref{figa1}, \ref{figa2}より、
%伝播モードは長波長でAlfv\'en wave、短波長で光になり、$- \mathrm{Im}~\omega$は$k$の単調増加関数であるため
%$\omega_I = - 2 \pi / \Delta t$の場合の特性速度を求めれば、減衰せずに伝播するモードの伝播速度の最大速度
%が求まることになる。

For this reason, 
we substitute $- 2 \pi / \Delta t$ into $\omega_I$ of Eq. (\ref{eq10}), 
and solve it with respect to $k^2$: 
%そこで上の$k$の４次方程式の$\omega_I$に$- \mathrm{Im}~\omega$を代入して$k^2$を求める。
\begin{equation}
k^2 = (\beta_1 + \sqrt{\beta_2} ) / \beta_3
\label{eq11}
\end{equation}
\begin{eqnarray}
\beta_1 &=& B^4 (B^x)^2 \sigma^4 - B^6 \sigma^3 (\sigma + 2 \omega_I) 
\nonumber
\\
&-& 2 B^4 \rho h \sigma^2 (2 \sigma^2 + 7 \sigma \omega_I + 6 \omega_I^2) 
+ \rho^2 h^2 (\sigma + 2 \omega_I)^2 [- 4 (B^x)^2 \sigma^2 + 2 \rho h (\sigma + 2 \omega_I)^2] 
\nonumber
\\
&-& B^2 \rho h \sigma (\sigma + 2 \omega_I) [-4 (B^x)^2 \sigma^2 + \rho h (5 \sigma^2 + 18 \sigma \omega_I + 16 \omega_I^2) ]
\\
\beta_2 &=& -8 \rho^2 h^2 \omega_I (\sigma + 2 \omega_I)^3 
    [B^2 \sigma + 2 \rho h (\sigma + \omega_I) ] [B^2 \sigma + \rho h (\sigma + 2 \omega_I) ]^3 
\nonumber
\\
  &+& [-B^6 \sigma^3 (\sigma + 2 \omega_I) + 2 \rho^2 h^2 (\sigma + 2 \omega_I)^2 
    \{2 (B^x)^2 \sigma^2 - \rho h (\sigma + 2 \omega_I)^2\} 
\nonumber
\\
&+& B^4 \sigma^2 \{(B^x)^2 \sigma^2 - 2 \rho h (2 \sigma^2 + 7 \sigma \omega_I + 6 \omega_I^2)\}
\nonumber
\\
 &+& B^2 \rho h \sigma (\sigma + 2 \omega_I) 
\{4 (B^x)^2 \sigma^2 - \rho h (5 \sigma^2 + 18 \sigma \omega_I + 16 \omega_I^2)\}]^2
,
\\
\beta_3 &=& \rho^2 h^2 (\sigma + 2 \omega_I) [B^2 \sigma + \rho h (\sigma + 2 \omega_I) ]
,
\\
\omega_I &=& - \frac{2 \pi}{\Delta t}
\end{eqnarray}
where Eq. (\ref{eq10}) has two solutions of $k^2$, 
and we adopt the larger one 
since $k$ is a real number. 
%$k^2$は２つの解を持つが、$k$は実数なので大きい方の解を採用している。

Substituting above $k^2$ into Eq. (\ref{eq9}), 
one can obtain the desired characteristic velocity. 
Since what we need is appropriate MHD characteristic velocity, 
%Since what we need is the fast magnetosonic velocity, 
%Since what we need is the Alfv\'en velocity in the laboratory frame, 
obtained velocity cannot be used as the characteristic velocity. 
However, 
if obtained velocity is not Alfv\'en velocity, 
it shows that 
we should use light velocity for the characteristic velocity. 
%the MHD approximation becomes bad. 
%In this case, 
%we use the speed of light as the characteristic velocity; 
%and otherwise we use the fast magnetosonic velocity of RMHD. 
%and otherwise we use the Alfv\'en velocity of RMHD in general coordinate. 
%得られた$k^2$をEq. (\ref{eq9})に代入すれば求める伝播速度が求まる。
%私たちが必要なのはfast magnetosonicの速度なので求まった伝播速度はそのままは使えないが、
%求まった速度が光速になった場合はMHD近似が悪くなっていると考えられるため、
%この場合は伝播速度を光速にし、それ以外の場合はRMHDのfast magnetosonic waveの伝播速度を
%使うことにする。

This method requires some further explanation. 
%なおこの方法にはいくつか注意が必要である。

First, 
%above dispersion relation Eq. (\ref{eq6}) is obtained 
%in the fluid comoving frame, 
%and one can not use in general frame. 
%Since the dispersion relation in general frame is so complicated 
%that we transform physical variables into those in the comoving frame in each mesh. 
note that 
the above method needs $B^2$, $B^x$, and $\Delta t^n$ in the comoving frame, 
and one should transform numerical data from Lab frame to comoving frame. 
%so only we have to do is calculate them in each mesh.  
%まず求めた分散関係は流体のcomoving frameでの解析であるため、
%そのままの値を使うことは出来ない。
%一般座標系の分散関係は非常に複雑であるため、私たちは各メッシュで物理量をcomoving frameの量に
%変換して計算している。
%この場合必要なのは、comoving frameの$B^2$, $B^x$, $dt$なので、これらのみを計算すれば良い。

%Next, 
%since the dispersion relation Eq. (\ref{eq6}) is for Alfv\'en mode, 
%$\omega_R$ becomes $0$ 
%when $B^x = 0$ 
%and $k / \sigma \ll 1$. 
%The numerical experiments implies that 
%$\omega_R^2$ becomes negative in this case, 
%so we use the fast magnetosonic velocity as the characteristic velocity 
%%so we use the Alfv\'en velocity as the characteristic velocity 
%when $\omega_R^2 < 0$ and $k / \sigma \ll 1$ 
%and the speed of light otherwise. 
%%次に、分散関係はAlfv\'en modeのものであるので$B^x = 0$では光速に帰着する短波長、low $\sigma$の
%%場合以外では伝播速度は$0$になる。
%%この場合は数値実験により求まった$\omega_R^2$が負になることが示唆されているため、
%%この時はMHDのfast magnetosonic waveの伝播速度を、
%%それ以外では光速を用いるようにしている。

Second, 
numerical experiments indicate that 
$\omega_R$ becomes $0$ for some range of $k$ 
for some parameter, 
and $k^2$ of Eq. (\ref{eq11}) becomes negative. 
In this case, 
we use speed of light as the characteristic velocity. 
%またパラメータによってはFig. \ref{figa1}, \ref{figa2}とは違い、
%ある範囲の波数$k$でRe $\omega$が$0$になることが数値実験により示唆されている。
%この場合、Eq. (\ref{eq11})の$k^2$が負になるため、この場合は伝播速度を
%光速にしている。

Finally, 
Fig. \ref{figa1} implies that 
$- \omega_I / \sigma$ has some maximum value. 
This can be proved as follows. 
First, 
dividing Eq. (\ref{eq6}) by $\sigma^4$, 
one obtains 
%最後にFig. \ref{figa2}から示唆されるように、$- \mathrm{Im}~\omega / \sigma$には上限がある。
%これは次のようにして示される。
%まずEq. (\ref{eq6})の両辺を$\sigma^4$で割り、$\bar{\omega} \equiv \omega / \sigma$, 
%$\bar{k} \equiv k / \sigma$とする。
\begin{equation}
\rho h \bar{\omega}^4
+ i (B^2 + 2 \rho h) \bar{\omega}^3 
- [\bar{k}^2 \rho h + (B^2 + \rho h) ] \bar{\omega}^2 
- i (B^2 + \rho h) \bar{k}^2 \bar{\omega} 
+ (B^x)^2 \bar{k}^2 = 0
,  
\label{eq12}
\end{equation}
where $\bar{\omega} \equiv \omega / \sigma$, and $\bar{k} \equiv k / \sigma$.

Fig. \ref{figa1} implies that 
the maximum value of $- \omega_I / \sigma$ is obtained in the limit of large $\bar{k}$, 
and the propagation mode is light wave in this limit. 
For this reason, 
we substitute $\bar{\omega} = \bar{k} - i \bar{\omega_I}'$ into Eq. (\ref{eq12}). 
Then it reduces to 
%Fig. \ref{figa2}より上限値は$\bar{k}$が無限大の極限であり、またこのとき伝播モードは光に
%なっているため、Eq. (eq:eq12)に$\bar{\omega} = \bar{k} - i \bar{\omega_I}'$を代入する。
\begin{eqnarray}
&-& i \rho (- 1 + 2 \bar{\omega_I}') \bar{k}^3
+ [(B^x)^2 + B^2 (- 1 + 2 \bar{\omega_I}') + \rho (- 1 + 5 \bar{\omega_I}' - 5 \bar{\omega_I}'^2) ] \bar{k}^2
\nonumber
\\
&+& i [- B^2 \bar{\omega_I}' (- 2 + 3 \bar{\omega_I}') 
  + 2 \rho \bar{\omega_I}' (1 - 3 \bar{\omega_I}' + 2 \bar{\omega_I}'^2) ] \bar{k}
\nonumber
\\
&-& B^2 (- 1 + \bar{\omega_I}') \bar{\omega_I}'^2 + \rho (- 1 + \bar{\omega_I}')^2 \bar{\omega_I}'^2 
= 0
. 
\end{eqnarray}
Since this is in the limit of large $\bar{k}$, 
what we have to consider is only the highest degree of $\bar{k}$. 
Then, 
we set its coefficient equal to $0$, 
and it reduces to 
%$\bar{k}$が無限の極限の場合、上式の$\bar{k}$の最高次のみを考えればよく、その係数を$0$とおくと
\begin{equation}
\bar{\omega_I}' = \frac{1}{2}
.
\end{equation}
%が得られる。
This shows that 
$- \omega_I / \sigma$ becomes $1 / 2$ in the limit of large $k / \sigma$, 
and we use the speed of light as the characteristic velocity 
when $\omega_I = - 2 \pi / \Delta t$ is less than $- \sigma / 2$. 
%よって$- \mathrm{Im} \omega / \sigma$は$1/2$が上限であることが分かった。
%そこで私たちは、$\omega_I = - 2 \pi / \Delta t$とした時に$- 1 / 2$を下回った場合は特性速度として光速を使うことにしている。

%\subsection{\label{sec:seca1}MUSCL}
\section{MUSCL}\label{sec:seca1}
%１次以上の精度にするにはRiemann solverや特性曲線法で数値流速を求める際に
%cell boundaryの物理量を精度の次数に従い計算する必要がある。
%この章ではVan LeerによるMUSCLを用いた２次精度の計算法について説明する。
For the second-order scheme, 
one has to compute the cell boundary numerical flux 
using Riemann solver or method of characteristics 
with left and right states obtained by using MUSCL. 
In this section, 
we explain MUSCL of Van Leer~\citep{V79}. 

%まず２次精度であるため、
%cell boundaryの物理量は次のようになる。
Since we need a second-order scheme, 
the left and right states of primitive variables $Q$ are 
\begin{eqnarray}
Q^{n+1/2}_{i+1/2,R} &=& Q^{n+1/2}_i + \frac{\delta Q^n_i}{2}
\label{eq01}
,
\\
Q^{n+1/2}_{i+1/2,L} &=& Q^{n+1/2}_{i+1} - \frac{\delta Q^n_{i+1}}{2}
\label{eq02}
,
\end{eqnarray}
where $Q^{n+1/2}$ follows from a predictor step 
%ここで$Q^{n+1/2}$は次のようなpredictor stepにより求まる。
\begin{equation}
U^{n+1/2}_i = U^n_i - \frac{\Delta t^n}{2 \Delta x_i} [F(Q^n_{i+1/2,L}) - F(Q^n_{i-1/2,R})]
,
\end{equation}
where $U$ is the conserved variables. 
In the above equation, 
$Q^{n}_{i \pm 1/2}$ can be computed from Eqs. (\ref{eq01}) and (\ref{eq02}) 
by replacing $Q^{n+1/2}$ with $Q^n$. 
%この式で$Q^n_{i \pm 1/2}$はEqs. \ref{eq01} and \ref{eq02}の$Q^{n+1/2}$を$Q^n$に
%置き換えて求めればよい。

When one uses MUSCL, 
the $\delta Q_i$ in Eqs. (\ref{eq01}) and (\ref{eq02}) are computed as follows: 
%MUSCLの場合、$\delta Q_i$は次のようにして求める。
\begin{equation}
(\delta Q_i)_{\mathrm{mono}} 
= 
\left\{
  \begin{array}{l}
    \mathrm{min} \left( 2 |\Delta Q_{i+1/2}|, |\Delta Q_i|, 2 |\Delta Q_{i-1/2}| \right) \mathrm{sgn} \Delta Q_i 
    \\
    \qquad \mathrm{if} \quad \mathrm{sgn} \Delta Q_{i+1/2} = \mathrm{sgn} \Delta Q_i = \mathrm{sgn} \Delta Q_{i-1/2}
    ,
    \\
    0 \qquad \mathrm{otherwise}
    ,
  \end{array}
\right.
\end{equation}
where
\begin{eqnarray}
\Delta Q_{i+1/2} = Q_{i+1} - Q_{i}
,
\\
\Delta Q_i = \frac{Q_{i+1} - Q_{i-1}}{2}
.
\end{eqnarray}
%以上の操作により時間、空間の２次精度が得られる。

\end{document}